\documentclass[twocolumn]{aastex61}
\pdfoutput=1 
\usepackage{amsmath,amstext}
\usepackage[T1]{fontenc}
\usepackage{apjfonts} 
\usepackage[figure,figure*]{hypcap}
\usepackage{natbib}
\usepackage{color}

\shorttitle{Fossilized Resonant Dropouts}
\shortauthors{Lawler et al.}

\begin{document}

\title{OSSOS: XIII. Fossilized Resonant Dropouts Tentatively Confirm Neptune's Migration was Grainy and Slow}

\author[0000-0001-5368-386X]{S.~M. Lawler}
\affiliation{Herzberg Astronomy and Astrophysics Research Centre, National Research Council of Canada, 5071 West Saanich Rd, Victoria, British Columbia V9E 2E7, Canada}
\author[0000-0003-4797-5262]{R.~E. Pike}
\affiliation{Institute of Astronomy and Astrophysics, Academia Sinica; 11F of AS/NTU Astronomy-Mathematics Building, No. 1 Roosevelt Rd., Sec. 4, Taipei 10617, Taiwan}
\author{N.~Kaib}
\affiliation{HL Dodge Department of Physics \& Astronomy, University of Oklahoma, Norman, OK 73019, USA}
\author[0000-0003-4143-8589]{M.~Alexandersen} 
\affiliation{Institute of Astronomy and Astrophysics, Academia Sinica; 11F of AS/NTU Astronomy-Mathematics Building, No. 1 Roosevelt Rd., Sec. 4, Taipei 10617, Taiwan}
\author[0000-0003-3257-4490]{M.~T. Bannister}
\affiliation{Astrophysics Research Centre, School of Mathematics and Physics, Queen's University Belfast, Belfast BT7 1NN, United Kingdom}
\author[0000-0001-7244-6069]{Y.-T. Chen} 
\affiliation{Institute of Astronomy and Astrophysics, Academia Sinica; 11F of AS/NTU Astronomy-Mathematics Building, No. 1 Roosevelt Rd., Sec. 4, Taipei 10617, Taiwan}
\author{B.~Gladman}
\affiliation{Department of Physics and Astronomy, University of British Columbia, Vancouver, BC V6T 1Z1, Canada}
\author{S.~Gwyn}
\affiliation{Herzberg Astronomy and Astrophysics Research Centre, National Research Council of Canada, 5071 West Saanich Rd, Victoria, British Columbia V9E 2E7, Canada}
\author[0000-0001-7032-5255]{J.~J. Kavelaars}
\affiliation{Herzberg Astronomy and Astrophysics Research Centre, National Research Council of Canada, 5071 West Saanich Rd, Victoria, British Columbia V9E 2E7, Canada}
\author[0000-0003-0407-2266]{J.-M. Petit}
\affiliation{Institut UTINAM UMR6213, CNRS, Univ. Bourgogne Franche-Comt\'e, OSU Theta F25000 Besan\c{c}on, France}
\author[0000-0001-8736-236X]{K.~Volk}
\affiliation{Lunar and Planetary Laboratory, University of Arizona, 1629 E University Blvd, Tucson, AZ 85721, USA}

\begin{abstract}

The migration of Neptune's resonances through the proto-Kuiper belt has been imprinted in the distribution of small bodies in the outer Solar System.  
Here we analyze five published Neptune migration models in detail, focusing on the high pericenter distance (high-$q$) trans-Neptunian Objects (TNOs) near Neptune's 5:2 and 3:1 mean-motion resonances, because they have large resonant populations, are outside the main classical belt, and are relatively isolated from other strong resonances.
We compare the observationally biased output from these dynamical models with the detected TNOs from the Outer Solar System Origins Survey, via its Survey Simulator.
All of the four new OSSOS detections of high-$q$ non-resonant TNOs are on the Sunward side of the 5:2 and 3:1 resonances. 
We show that even after accounting for observation biases, this asymmetric distribution cannot be drawn from a uniform distribution of TNOs at 2$\sigma$ confidence.
As shown by previous work, our analysis here tentatively confirms that the dynamical model that uses grainy slow Neptune migration provides the best match to the real high-$q$ TNO orbital data. 
However, due to extreme observational biases, we have very few high-$q$ TNO discoveries with which to statistically constrain the models.
Thus, this analysis provides a framework for future comparison between the output from detailed, dynamically classified Neptune migration simulations and the TNO discoveries from future well-characterized surveys.
We show that a deeper survey (to a limiting $r$-magnitude of 26.0) with a similar survey area to OSSOS could statistically distinguish between these five Neptune migration models.

\end{abstract}


\section{Introduction}

The Kuiper Belt as observed today looks very different from the dynamically cold, flat disk of leftover planetesimals originally hypothesized to exist beyond Neptune's orbit \citep{Edgeworth1949}.
Several dynamically distinct components have been detected in the Kuiper belt, some of which are highly excited, i.e.\ on very eccentric, inclined orbits \citep[e.g.,][]{Gladmanetal2008}.  
The largest fraction of trans-Neptunian objects (TNOs) do indeed reside within the main classical belt on dynamically cold orbits.
The Kuiper Belt also contains a large fraction ($\sim$20\%) of TNOs on orbits that are in mean-motion resonances with Neptune \citep{Gladmanetal2012,Adamsetal2014}, and a significant scattering component that is on highly excited, dynamically unstable orbits \citep{Gladman2005,Shankmanetal2013}. 
Of interest in this work are the population of non-resonant TNOs with pericenters outside the gravitational scattering influence of Neptune: the ``detached'' population \citep{Gladmanetal2002}.

Increasingly detailed dynamical simulations over the years have shown that much of the observed orbital structure of the Kuiper Belt can be caused by the outward migration of Neptune's orbit, with incrementally powerful constraints on the exact timing and mode of Neptune's migration \citep[e.g.,][]{Malhotra1993,Thommesetal1999,Tsiganisetal2005,BrasserMorbidelli2013,NesvornyVokrouhlicky2016}.
In order to compare these detailed simulations to observations of the distribution of TNO orbits, it is vital that observational biases are understood and accounted for.  
Particularly for the high pericenter distance (high-$q$) TNOs, the observational biases are severe and can have unintuitive consequences for the detected orbital distributions \citep[for a detailed discussion of these observational bias effects, see][]{Shankmanetal2017}.
A thorough and well-tested way to account for observing biases is to carefully record a survey's depth, pointing direction and area on the sky, tracking fraction, and detection efficiency.  
These biases can then be applied to the output from a detailed dynamical simulation by Survey Simulator software, and the simulated detections can then be directly compared with the real survey detections in a statistically significant way, effectively ``debiasing'' the survey \citep{Jonesetal2006}. 
This technique can only be applied to TNOs discovered in surveys that carefully record their biases; TNOs pulled from the Minor Planet Center Database without regard for discovery survey cannot be debiased because most surveys do not publish their biases and survey characteristics.

In this paper we take the results from five recently published, detailed Neptune migration simulations and compare them to the detected TNOs from the Outer Solar System Origins Survey \citep[OSSOS;][]{Bannisteretal2016,Bannisteretal2018}, making use of the OSSOS Survey Simulator \citep{LawlerSurveySimulator,Petitetal2018}.
We particularly focus on the high-pericenter TNOs as a powerful ``fossilized'' tracer of Neptune's migration.  
In Section~\ref{sec:highq} we discuss the dynamics of high-$q$ TNOs, as well as review in detail the properties of the migration simulations that we analyze (Section~\ref{sec:sims}) and the structure of the near-resonant orbital element distributions for the different models (Section~\ref{sec:5231}).
In Section~\ref{sec:comparing} we use the OSSOS Survey simulator (Section~\ref{sec:surveysim}) to apply the survey biases to the models, starting with a simple uniform distribution (Section~\ref{sec:uniform}), then observationally biasing the dynamical models (Section~\ref{sec:surveysimsetup}) and comparing them with the real OSSOS detections. 
While so few real detections provide little statistical constraint on the models, we conclude that overall, grainy slow Neptune migration provides the best match (Section~\ref{sec:gsbest}), and this analysis provides a framework for how to compare these future TNO detections with the dynamical model output.
We conclude with a comparison to the MPC database (Section~\ref{sec:mpc}) and a discussion of how resonant sticking may be important to explain high-$q$ TNOs on larger semimajor orbits (Section~\ref{sec:dropouts}).
In the Appendix, we provide detailed properties and analysis of a future, deeper survey that would be able to statistically distinguish between these migration models.

\section{High Pericenter TNOs as a Diagnostic of Neptune's Migration} \label{sec:highq}

\begin{figure}
\begin{center}
\includegraphics[scale=0.6]{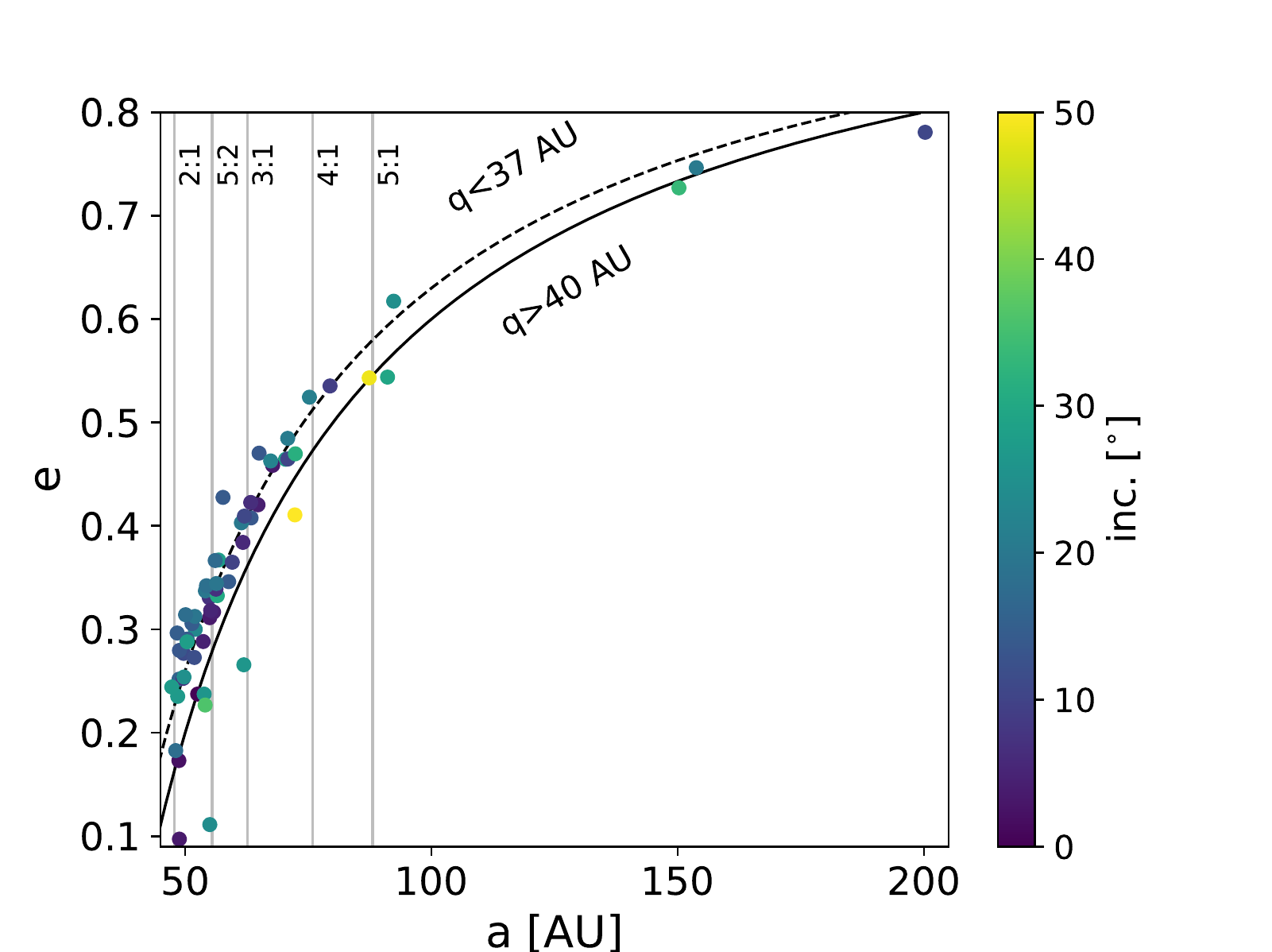}
\caption{
Orbital elements of the outer classical and detached TNOs detected by the OSSOS ensemble of surveys \citep{Bannisteretal2018}; these are the non-resonant TNOs at larger $a$ than the 2:1 resonance that are have stable orbits for at least 10~Myr.
Inclination is shown by color.
The solid curve shows $q=40$~AU and the dashed curve shows $q=37$~AU.
Points below the solid black curve are high pericenter TNOs ($q>40$~AU).
The semimajor axes of several well-populated resonances are shown with grey lines for reference; all detected high-$q$ TNOs in the OSSOS ensemble are within $\pm2$~AU of strong resonances, which is the focus of this work.
}
\label{fig:detached}
\end{center}
\end{figure}

TNOs with pericenters outside the gravitational scattering influence of Neptune either formed there (as did the classical Kuiper belt on fairly circular orbits) or must have been emplaced later by some other interaction.
TNOs that are not resonant and are on orbits that remain stable for $>$10~Myr timescales are either part of the classical belt or part of the ``detached'' TNO population, as they are currently dynamically decoupled from Neptune's scattering influence, unlike the ``scattering'' TNOs and Centaurs which experience dynamical instability on $<$10~Myr timescales \citep{Gladmanetal2008}.
The border between the ``scattering'' and ``detached'' TNOs doesn't correspond to a simple cut in pericenter distance $q$, but $q>37$~AU is often used to classify which TNOs are not currently experiencing strong gravitational encounters with Neptune \citep[e.g.][]{LykawkaMukai2007}.
For our purposes, we define both the outer classicals and the detached TNOs as part of the distant dynamically detached population. The outer classical TNOs (defined in \citealt{Gladmanetal2008} as $>$10~Myr stable orbits having $e<0.24$ and $a>47.8$~AU, the location of the 2:1 mean-motion resonance with Neptune), and the detached TNOs (defined in \citealt{Gladmanetal2008} as $>$10~Myr stable orbits having $e>0.24$ and $a>47.8$~AU).  
Figure~\ref{fig:detached} shows the orbital properties of all of the detached and outer classical TNOs that were detected by the OSSOS ensemble of surveys (see Section~\ref{sec:surveysim}).

Without a mechanism to create high-$q$ detached TNOs, there was originally no expectation to discover eccentric high-$q$ TNOs.  
The first detections showed that a surprisingly large high-$q$ population was required due to the extreme observational biases against detecting these high-$q$ TNOs: their large pericenter distances mean that they always remain at large heliocentric distances and thus faint magnitudes \citep{Gladmanetal2002}.

\citet{Gomes2003} provided a possible explanation for these high-$q$ orbits, showing that Kozai cycling inside mean-motion resonances can simultaneously raise inclinations and pericenters.
These TNOs are then more likely to fall out of resonances, because the resonances are thinner at the extreme values of inclination and eccentricity.
The objects that fall out of resonance at the high-$e$ and low-$i$ end of a Kozai cycle are likely to be Neptune-crossing, and once outside the protection of a mean-motion resonance become part of the scattering population and are dynamically unstable on timescales of a few Myr.  
But the objects that fall out of resonance at the low-$e$ and high-$i$ end of a Kozai cycle are dynamically decoupled from Neptune due to their high pericenter distance $q$, and thus should remain in those orbits as a ``fossilized'' remnant of Neptune's past orbital migration history.
Due to their long-term stability, these high-$q$ resonant dropouts are the focus of this work.

\subsection{Dynamical Simulations of Neptune's Migration}
\label{sec:sims}

Neptune's migration was first invoked to explain Pluto's eccentric resonant orbit \citep{Malhotra1993}.
Since that first Neptune migration models, many migration models and subsequent variations have been proposed, resulting in gradually better fits to the orbital structure of the steadily increasing number of known TNOs.
The five models that we analyse here were chosen because they contain many particles and are dynamically classified according to the criteria laid out in \citet{Gladmanetal2008} \citep[as is the case for][]{PikeLawler2017}\footnote{Dynamically classified model from \citet{Pikeetal2017} is publicly available here: \url{https://doi.org/10.11570/16.0009}}, or because the authors provided their simulation output in a manner that was able to be dynamically classified through additional dynamical simulations \citep[as is the case for models from][]{KaibSheppard2016}\footnote{Dynamically classified models from \citet{KaibSheppard2016} are available here: \url{https://doi.org/10.11570/19.0008}}.

\citet{Pikeetal2017} and \citet[][hereafter referred to as {\bf PL17}]{PikeLawler2017} present a detailed analysis of the giant planet migration simulation from \citet{BrasserMorbidelli2013}.
The planetary orbital evolution path within this simulation is based on the ``Nice model'' simulation from \citet{Levisonetal2008}.
The Nice Model \citep{Thommesetal1999,Tsiganisetal2005} invokes dynamical instability and scattering between the giant planets, causing Neptune to attain high eccentricity (maximum $e\simeq0.3$) as it migrates outwards while damping to a more circular orbit by scattering planetesimals.
Neptune's high eccentricity phase in this model has been demonstrated to cause problems with the structure of the classical belt \citep{DawsonMurrayClay2012}, 
the resonant TNO populations \citep{Gladmanetal2012}, and the asteroid belt \citep{RoigNesvorny2015}.
Subsequent simulations show that limiting Neptune's maximum eccentricity and migration timescale can preserve the cold classical population \citep{Batyginetal2011,DawsonMurrayClay2012,Ribeiroetal2018}, but it is difficult to create this starting condition for Neptune.
However, the detailed structure is testable because of the large number of particles in this simulation, and if other aspects of the outer Solar System structure provide a compelling match to the real detections then it would be worthwhile to consider ways to include a high $e$ phase of Neptune in migration models.
It is also a simpler model in some ways than models that require an additional giant planet that is later ejected.
This simulation produces ``beards'' of objects that drop out of resonance on both sides of the resonance, caused by the combination of Kozai cycling and narrowing of the resonance as Neptune's orbit circularizes.

\citet{KaibSheppard2016}, hereafter referred to as {\bf KS16}, provide a detailed model of the outer Kuiper belt structure resulting from grainy or smooth migration combined with slow or fast migration timescales. 
These four KS16 models build upon previously published work, and agree with similar simulations in \citet{Nesvornyetal2016}.
In each of the KS16 simulations, Neptune is scattered but never attains a very high eccentricity ($e<0.1$), and the migrations all include a ``jump'' in $a$ and $e$ meant to simulate the scattering of a now-ejected ice giant \citep[as in][]{Nesvorny2015b}, sometimes referred to as the ``jumping Jupiter'' model \citep{Brasseretal2009,Morbidellietal2010}.
Two of the models include only smooth migration \citep{Malhotra1993,HahnMalhotra2005}, and two use ``grainy migration,'' first discussed in \citet{NesvornyVokrouhlicky2016} and further analysed in \citet{Nesvornyetal2016}.
In grainy migration, Neptune has small jumps in semimajor axis to simulate scattering of the largest planetesimals (roughly $\sim$1000~km, or Pluto-sized) and this affects capture and retention of TNOs in resonances and appears to produce an overall Kuiper belt structure that more closely matches observations \citep{NesvornyVokrouhlicky2016}.
More recent work \citep{ShannonDawson2018} uses binaries, the cold classical TNOs, and resonant TNOs to place a limit on the number of $\sim$1000~km planetesimals that could have existed prior to Neptune's migration, which agrees with the previous analysis.
The timescales for these migrations have $e$-folding timescales of 10~Myr pre-jump and 30~Myr post-jump for the \emph{fast migrations}, and 30~Myr pre-jump and 100~Myr post-jump for the \emph{slow migrations}.
These timescales shown to provide good matches to the inclination distribution of TNOs \citep{Nesvorny2015a}

To avoid confusion, we define TNOs closer to the Sun than a given resonance as ``\emph{Sunward}'' and those more distant as ``\emph{Outward}.''
Both \citet{Nesvornyetal2016} and KS16 show ``trails'' of particles that drop out of the resonances on the Sunward side as the resonances march outwards with Neptune's migration.  
KS16 show this effect is stronger in grainy migration than smooth migration, and stronger in slower migrations than faster migrations.

Throughout the paper, we use the following abbreviations to refer to the five simulations.
\begin{itemize}
\item All {\bf KS16} simulations keep Neptune at relatively low eccentricity ($e<0.1$), with a jump in Neptune's $a$ and $e$ that is caused by scattering of a now-ejected additional ice giant.
\begin{itemize}
\item {\bf KS16 GF}: Grainy migration with a fast timescale ($e$-folding timescale of 10~Myr and 30~Myr pre- and post-Neptune's large jump in $a$, respectively).
\item {\bf KS16 GS}: Grainy migration with a slow timescale ($e$-folding timescale of 30~Myr and 100~Myr pre- and post-Neptune's large jump in $a$, respectively).
\item {\bf KS16 SmF}: Smooth migration with the same fast timescale above.
\item {\bf KS16 SmS}: Smooth migration with the same slow timescale above.
\end{itemize}
\item {\bf PL17}: Neptune scatters to high eccentricity ($e\simeq0.3$) and circularizes while migrating outwards with a timescale of $\sim$100~Myr.
\end{itemize}

The general effect of Kozai cycling combined with Neptune's migration is different for each of the five simulations we analyse.
Most of the dropout appears to happen during Neptune's circularization phase in PL17, resulting in rather symmetric ``beards'' of near-resonant objects at high-$q$ (see Figure~1 in that work).
The four simulations in KS16 form ``trails'' of near-resonant, high-$q$ objects preferentially populating the Sunward side of resonances (see Figure~2 in that work). 
The trails are of varying properties and densities depending on the simulation.
Grainy migration appears to more easily drop objects out of resonance, as does slower timescale Neptune migration.

\subsection{The 5:2 and 3:1 Resonances} \label{sec:5231}

We focus on the 5:2 and 3:1 resonances in this analysis because they are powerful resonances, relatively isolated in semi-major axis, less contaminated by resonant sticking (unlike more Sunward resonances that are embedded in the classical Kuiper belt), and close enough that surveys have  a significant number of detected TNOs both inside the resonance and at nearby semimajor axes.
The OSSOS ensemble of surveys includes TNO detections around each of these resonances, so it is possible to make a statistical comparison of the real detections to the simulation results.
This requires a careful selection of real TNOs and a dynamical classification of the orbits of model objects to diagnose resonance and non-resonance (see Section~\ref{sec:surveysim}).

\begin{figure*}
\begin{center}
\includegraphics[scale=0.44]{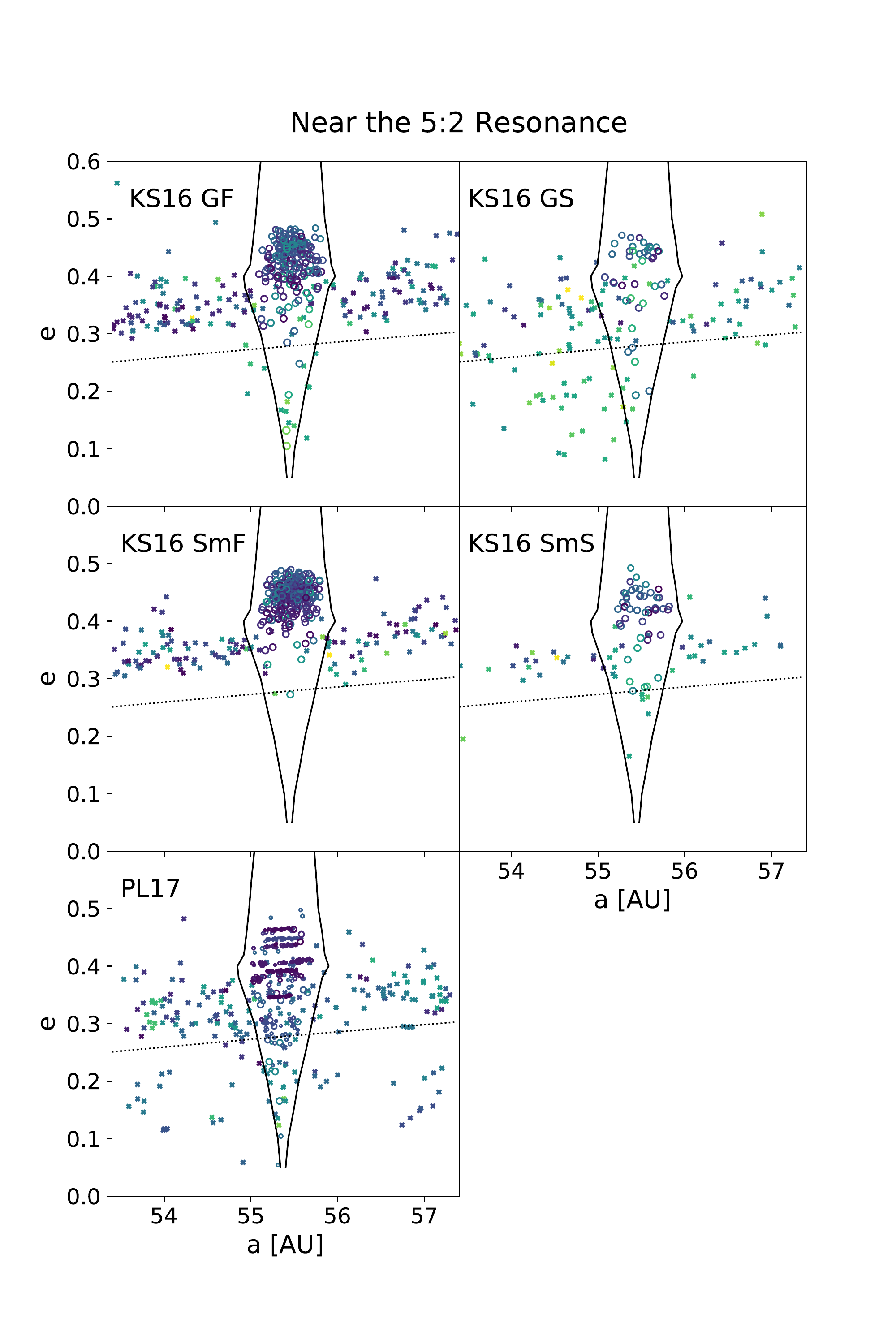} \includegraphics[scale=0.44]{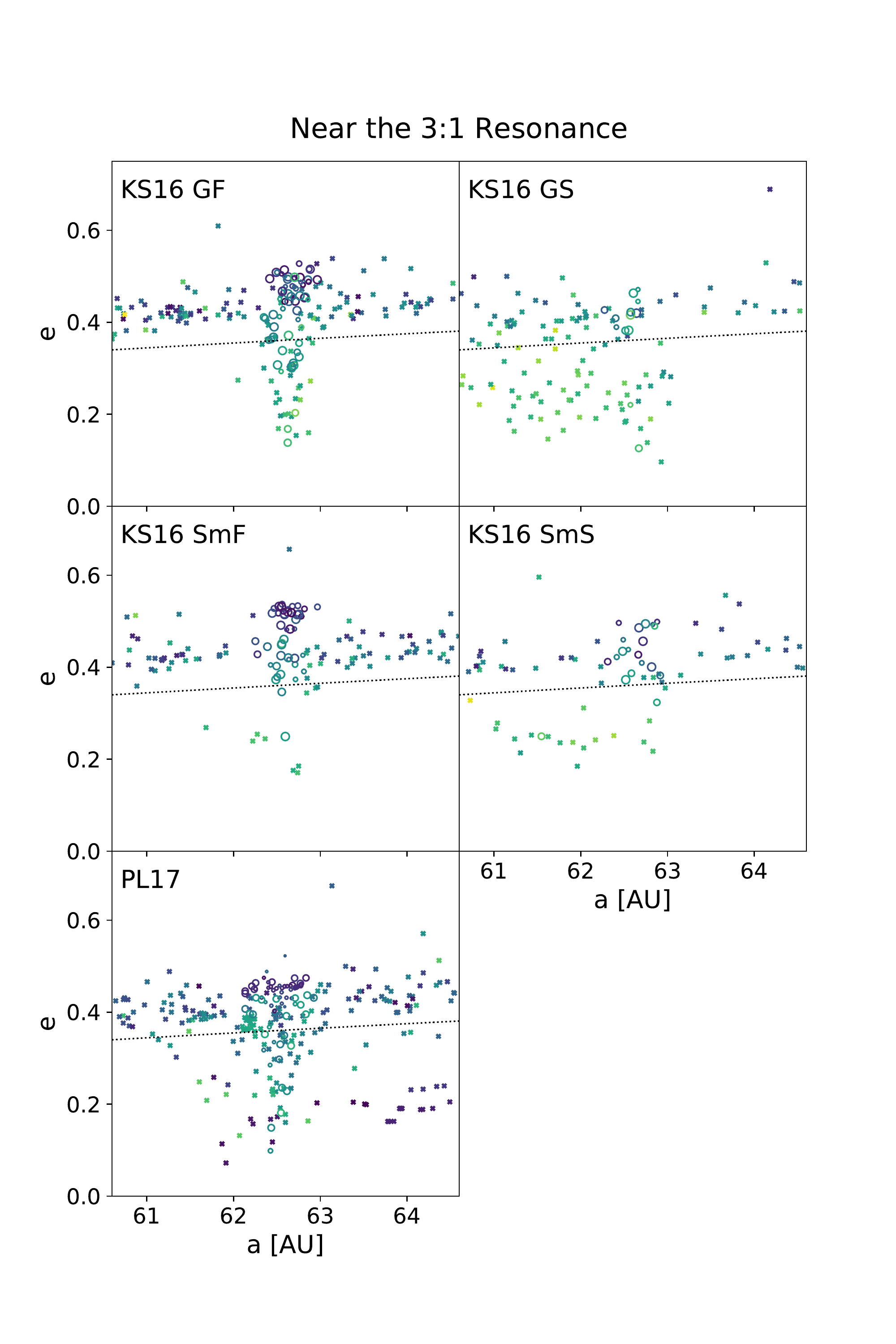}
\caption{
Comparison between the simulated objects with $a$ within $\pm2$~AU of the 5:2 ($a=55.4$~AU, left plots) and 3:1 resonance centers ($a=62.6$~AU, right plots). 
Circles show objects that we classify as resonant with circle size proportional to libration amplitude, x's show non-resonant objects, and color shows inclination (same color scale as Figure~\ref{fig:detached}).  
The objects located below the dotted black curve have $q>40$~AU and are the high-pericenter objects we use for comparison (Section~\ref{sec:comparing}).
The grey solid line (right plots) shows the 3:1 resonance center, while the solid black lines (left plots) show the 5:2 resonant stability limits computed in \citet{Malhotraetal2018} as a guide; our dynamical classification generally agrees with these limits.
}
\label{fig:near52}
\end{center}
\end{figure*}

The particles within the five simulations have been classified into dynamical classes using 10-30~Myr orbital integrations including the four giant planets to diagnose resonance occupation \citep[following the TNO classification scheme in][]{Gladmanetal2008}.  
The PL17 simulation particles are dynamically classified in \citet{Pikeetal2017}, while the KS16 particles are classified here.  
The orbital structure around the 5:2 and 3:1 resonances are shown for all five simulations in Figure~\ref{fig:near52} (5:2 in left panels, 3:1 in right panels).
In each of the plots, circles show particles that are resonant at the end of the simulation, with point size proportional to libration amplitude, and non-resonant particles shown with x's.  
Color shows inclination, with purple showing lowest inclinations ranging to yellow for highest inclinations ($\sim50^{\circ}$, same color scale as Figure~\ref{fig:detached}).  
The points below the black dotted curve are objects with $q>40$~AU.

Some general trends are worth discussing in the five simulations.  
The resonant dropouts (high-$q$ objects marked by x's) are almost entirely Sunward of the resonances in the two grainy KS16 simulations, are nearly absent in the two smooth KS16 simulations for the 5:2, but present Sunward for the 3:1, and are present on both Sunward and Outward sides of the resonances in the PL17 simulation.
The high-$q$ resonant dropouts in the grainy KS16 simulations are on average higher inclination that the lower-$q$ (higher-$e$) particles in the simulation (as shown by the yellow and green color rather than blue and purple).  
This supports the idea that the Kozai effect is important in these resonant dropouts: high-$q$ (and correspondingly low-$e$) particles have high inclinations when they fall out of resonance and enter the fossilized detached population.
The high inclination effect is not as obvious in the Nice model PL17 particles.
This is likely due to Neptune's initially high eccentricity ($e\simeq0.3$) causing very wide resonances in $a$.  
The combination of Neptune's orbital circularization causing the resonance boundaries to narrow and Kozai oscillations caused many objects to drop out of resonance on both the Sunward and Outward sides.

\begin{figure}
\begin{center}
\includegraphics[scale=0.45]{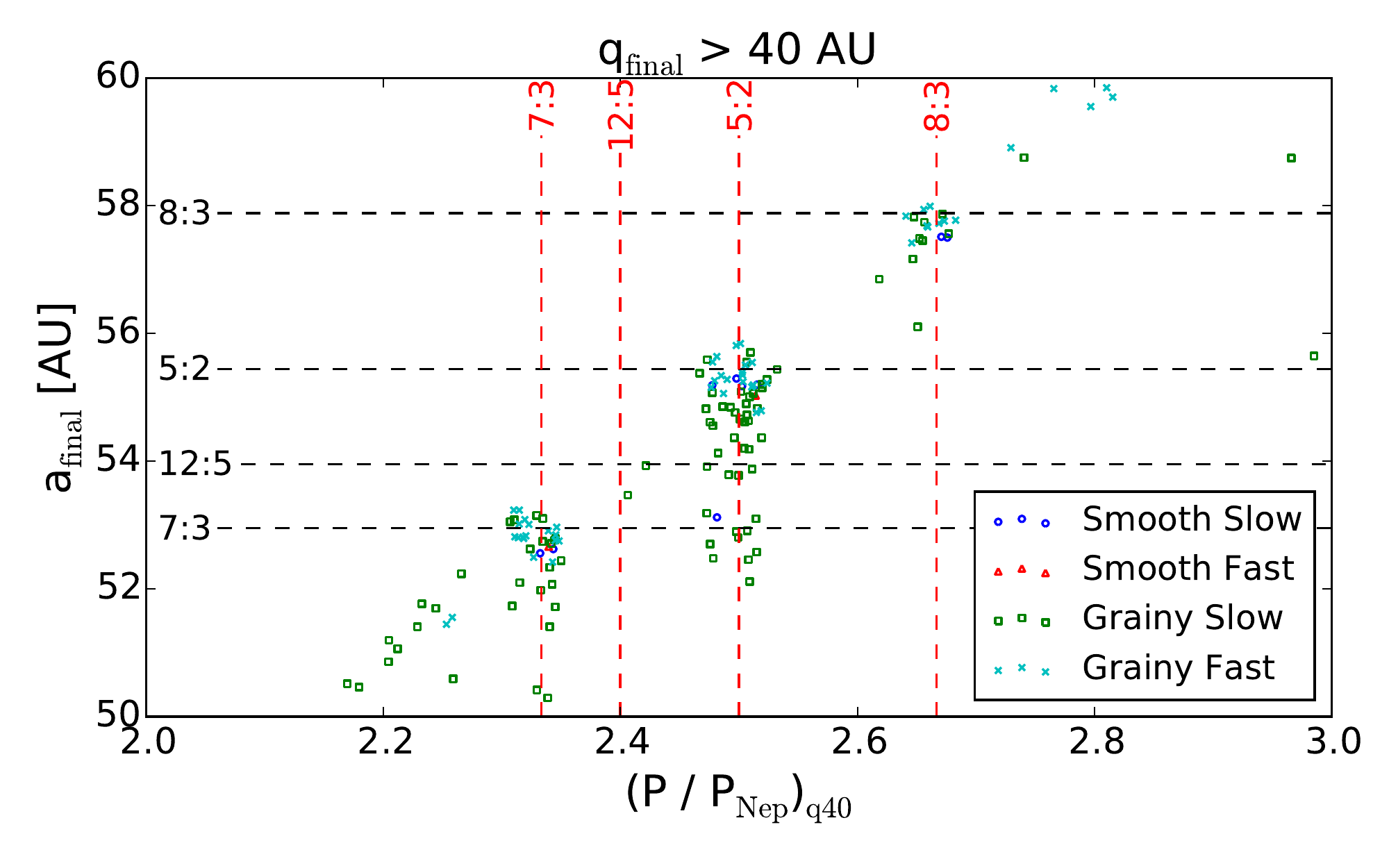}
\caption{
Every particle that ever attains high pericenter ($q>40$~AU) in the KS16 models and their instantaneous period ratio with Neptune at the timestep where the particle first attained $q>40$~AU (x-axis), compared with the object's final $a$ (y-axis).
The high-$q$ objects in all of the KS16 simulations that end the simulation with $a$ close to the 5:2 (those near the black horizontal dashed 5:2 line) first attain their high-$q$ almost exclusively inside the 5:2 (that is, they cluster along the red vertical dashed 5:2 line), so contamination by pericenter raising inside other nearby resonances is not a significant effect.
}
\label{fig:phist}
\end{center}
\end{figure}

There are several higher-order resonances close to the 5:2 resonance that each have TNOs observed to be currently occupying these resonances \citep[e.g., the OSSOS ensemble discovered TNOs occupying the 7:3, 12:5, 13:5, and 8:3 resonances, all with $a$ within $\pm2$~AU of the center of the 5:2 resonance;][]{Bannisteretal2018}.
There are no known TNOs within $\pm2$~AU of the 3:1 resonance that occupy a resonance other than the 3:1.
All of the known resonant TNOs within $\pm2$~AU of the 5:2 in resonances other than the 5:2 have $q<40$~AU, thus don't fall into our high-$q$ sample.  
For the high-$q$ particles created in the course of the Neptune migration simulations, we check that the pericenter-raising occurs primarily inside the 5:2 and not inside the nearby resonances by checking resonance occupation at the timestep where a particle first attains $q>40$~AU (shown in Figure~\ref{fig:phist}).
The number of particles with period ratios close to the 5:2 when attaining $q>40$~AU exceeds the number close to the other resonances (particularly the 12:5, which is closest) by a factor of several.  

The next step is to compare these models with the real detected TNOs.
However, a direct comparison between the models and the detected TNOs would be misleading because of the severe detection biases based on distance and survey characteristics.
We are able to compare the carefully selected near-resonant test particles with the real TNOs near resonance by using the OSSOS Survey Simulator.

\section{Accounting for Observational Biases: Comparing Models to Observations} \label{sec:comparing}

\subsection{The Well-Characterized OSSOS Survey} \label{sec:surveysim}

The Outer Solar System Origins Survey \citep[OSSOS;][]{Bannisteretal2016,Bannisteretal2018} was a large program on the Canada-France-Hawaii Telescope over five years.
It is a well-characterized survey, where all of the survey block pointings, the detection efficiencies, and tracking efficiencies and rate cuts at different magnitude limits are well known and published, so all of these biases can be applied to a dynamical model output, and the forward-biased model can be directly statistically compared to the detected TNOs in the survey \citep{LawlerSurveySimulator}.
The OSSOS Survey, in combination with three other well-characterized surveys \citep{Petitetal2011,Alexandersenetal2016,Petitetal2017} is referred to as the ``OSSOS ensemble'' of surveys, and in total detected 1142 TNOs with exceptionally well-measured orbits.
This Survey Simulator\footnote{The OSSOS Survey Simulator software is publicly available at \url{https://github.com/OSSOS/SurveySimulator}.} methodology has previously been used to constrain the size distribution and population of scattering TNOs \citep{Shankmanetal2013,Shankmanetal2016,LawlerScattering}, 
the populations and orbital structure within different mean-motion resonances \citep{Gladmanetal2012,LawlerKozai,Pikeetal2015,Alexandersenetal2016,Volketal2016,Volketal2018}, 
and the populations and size distributions of the main classical belt \citep{Petitetal2011}.
The Survey Simulator methodology has also been used to compare the output from dynamical simulations with observed TNO distributions, providing support for grainy migration \citep{NesvornyVokrouhlicky2016}, making predictions about observations to test the Nice model \citep{Pikeetal2017}, and constraints on the the population, orbital distribution, and mass of the distant Kuiper Belt in the presence of an undiscovered giant planet \citep{Lawleretal2017,Shankmanetal2017P9,LawlerSurveySimulator}.

This survey characterization is critical for quantifying selection effects that might affect the interpretation of OSSOS detections.
The biases against detection of high-$q$ TNOs are extreme and unintuitive \citep{Shankmanetal2017}, so using a Survey Simulator to test and account for these biases is vital.  
The near-resonant TNOs do not have a longitudinal bias, but the Sunward TNOs are slightly more detectable than the Outward TNOs because of their smaller semi-major axes.
The Survey Simulator and characterized survey are crucial to understanding the significance of this effect, the relative detectability of nearby resonant TNOs, and to interpreting the detected Sunward/Outward fractions.

\begin{figure}
\begin{center}
\includegraphics[scale=0.45]{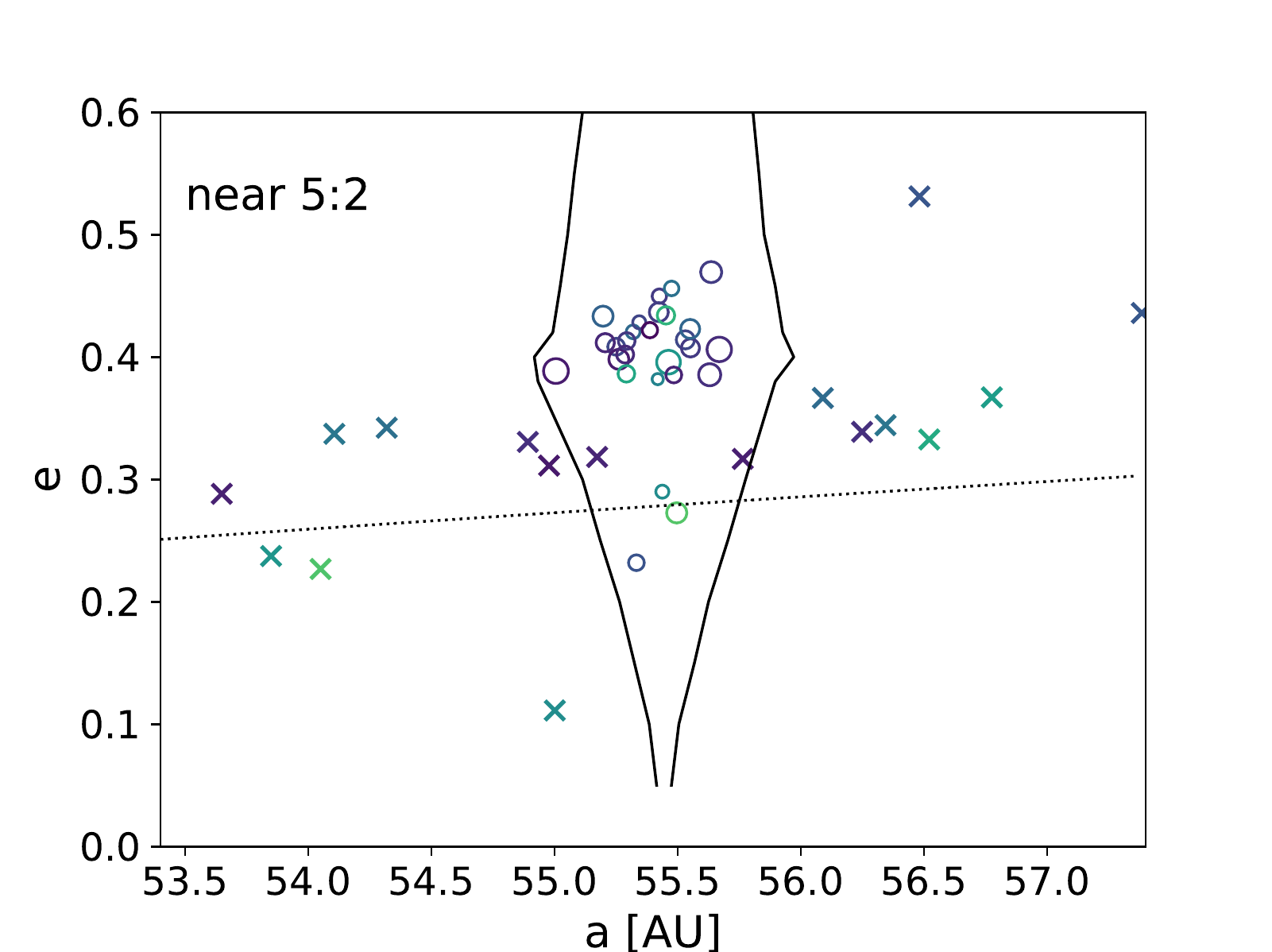}
\includegraphics[scale=0.45]{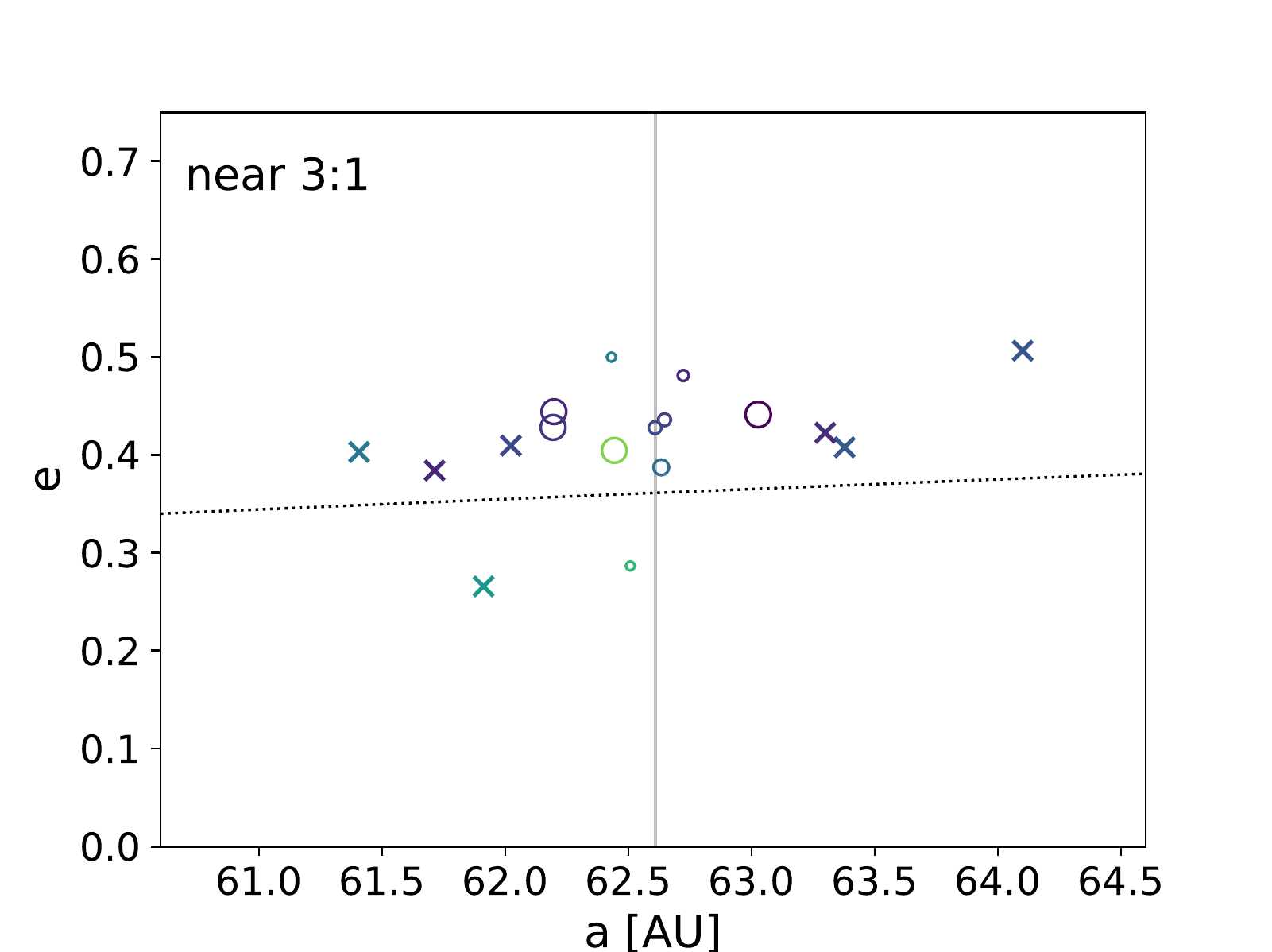}
\caption{
All the TNOs detected by OSSOS within $\pm$2AU of the 5:2 and 3:1 resonances (top and bottom panels, respectively).
TNOs that are not in the 5:2 or 3:1 resonances are indicated by `x' symbols, while circles show 5:2 and 3:1 resonators, with larger circle sizes showing larger libration amplitudes.  
Color indicates inclination, as in Figure~\ref{fig:detached}.
}
\label{fig:ossos}
\end{center}
\end{figure}

We focus on the OSSOS TNO detections within $\pm$2~AU of the 5:2 and 3:1 resonance, shown in Figure~\ref{fig:ossos}.
TNOs in this $a$-range that are not in the 5:2 or 3:1 resonances are indicated by an `x,' while TNOs in the 5:2 and 3:1 resonances are shown by circles, with larger circles indicating larger libration amplitudes within the resonance.

\subsection{Sunward vs.\ Outward Near-Resonant High-$q$ Objects: Not A Uniform Distribution} \label{sec:uniform}

The OSSOS survey discovered a number of near-resonant and high-$q$ objects.
We focus on the 5:2 and 3:1 resonance because these resonances are relatively isolated, and as demonstrated in Section \ref{sec:5231}, these resonances are responsible for the vast majority of the high-$q$ TNOs fossilized within $\pm2$~AU of those resonances.
The OSSOS ensemble \citep{Bannisteretal2018} detected 27 TNOs currently librating in the 5:2 resonance, and 11 in the 3:1 resonance.
Within $\pm2$~AU of the 5:2 resonance, OSSOS detected three non-resonant TNOs with $q>40$~AU, all Sunward of the resonance: {\tt L5c16} (2005~CG$_{81}$), {\tt HL7j5} (2007~LE$_{38}$), and {\tt o5m86} (2015~KQ$_{174}$).
Within $\pm2$~AU of the 3:1 resonance, OSSOS detected one non-resonant TNO with $q>40$~AU, also Sunward of the resonance: {\tt o3l72} (2013~SK$_{100}$).
These high-$q$ TNOs are listed in Table~\ref{tab:highq}.
We test the significance of the discovery of all four near-resonant objects Sunward of the resonances by using the OSSOS Survey Simulator to account for discovery biases.

\begin{deluxetable*}{ll|cccccccc}
\tabletypesize{\small}
\tablecolumns{9} 
\tablewidth{0pt}
\tablecaption{High-$q$ near-resonant TNOs detected by the OSSOS ensemble \label{tab:highq}}
\tablehead{ \colhead{OSSOS} & \colhead{MPC} & \colhead{$a$} & \colhead{$e$} & \colhead{$i$} & \colhead{$q$} & \colhead{$\Omega$}  & \colhead{$\omega$} & \colhead{T peri} & \colhead{$H_g$} \\
\colhead{name} & \colhead{designation} & \colhead{[AU]} & & \colhead{[$^{\circ}$]} & \colhead{[AU]} & \colhead{[$^{\circ}$]} & \colhead{[$^{\circ}$]} & \colhead{[JD-2400000]} & \colhead{[mag]}
 } 
\startdata
\multicolumn{2}{c}{\underline{5:2 resonance:}} & 55.4$^a$ & & & & & & \\
L5c16	&	2005~CG$_{81}$	&	53.849	&	0.23735	&	26.153	&	41.068	&	134.664	&	56.94	&	67197	&	6.14$^b$ \\
HL7j5	&	2007~LE$_{38}$	&	54.050	&	0.2268	&	35.966	&	41.794	&	193.529	&	53.67	&	53730	&	6.93 \\
o5m86	&	2015~KQ$_{174}$	&	55.00	&	0.111	&	24.398	&	48.89	&	213.97	&	290.20	&	25373	&	7.27 \\ \hline
\multicolumn{2}{c}{\underline{3:1 resonance:}} & 62.6$^a$ & & & & & & \\
o3l72	&	2013~SK$_{100}$	&	61.912	&	0.26569	&	26.288	&	45.462	&	17.112	&	11.76	&	61191	&	7.49 \\
\enddata
\tablenotetext{a}{For reference, we give $a$ for the 5:2 and 3:1 resonances from $a_{\rm N}(p/q)^{2/3}$}
\tablenotetext{b}{$H_r$ has been converted to $H_g$ using the relation $g-r=0.7$; see \citet{Shankmanetal2016}}
\tablecomments{For errors on measured orbital elements, see \citet{Bannisteretal2018}. All digits given are significant.}
\end{deluxetable*}

\begin{figure}
\begin{center}
\includegraphics[scale=0.45]{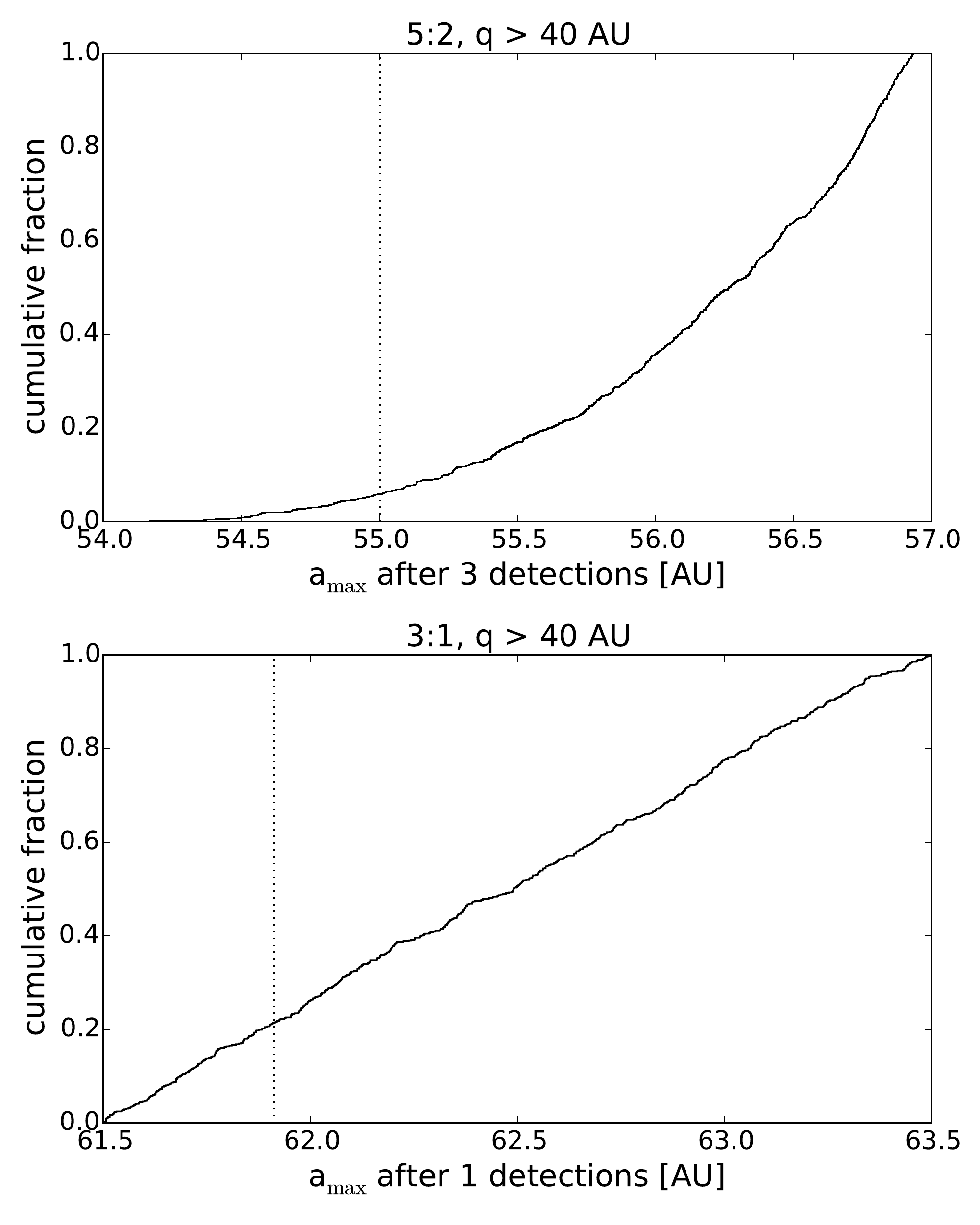}
\caption{
The largest-$a$ objects detected from a uniform period ratio distribution, for three detections near the 5:2 resonance (top plot, resonance center at $a=55.4$~AU) and 1 detection near the 3:1 resonance (bottom plot, resonance center at $a=62.6$~AU).  A distribution of high-$q$ TNOs drawn from a uniform distribution of period ratios was supplied to the Survey Simulator, which was run until the same number of detections as the OSSOS Ensemble surveys were found (three near-5:2 objects and one near-3:1 object).  The largest $a$ of the \emph{real} detections near each resonance is shown by the dashed line.  Using the OSSOS Survey Simulator, we find that the likelihood that the underlying distribution is uniform is 5.9\% for the 5:2 and 20\% for the 3:1, so the likelihood of both distributions finding low-$a$ members is 2.9\%.
}
\label{fig:amax}
\end{center}
\end{figure}

Comparing the OSSOS detections to a uniform distribution around the resonances indicates that the intrinsic distribution is significantly non-uniform.
We created orbital distributions of high-$q$ objects around the 5:2 and 3:1 MMR.
We used the combined objects from all {\bf KS16} models with $q>40$ AU and period ratios of 2.4-2.6$~P_{N}$ and 2.9-3.1~$P_N$ for the 5:2 and 3:1, respectively. Using the $q$- and $i$-distributions of these objects, we then randomly sampled from the synthetic objects assuming a uniform distribution in this range of period ratios.
Orbital angles were randomly assigned, and the synthetic objects were input into the survey simulator.
The near-5:2 simulations ran until 3 objects were detected and the near-3:1 simulations ran until just 1 object was detected.
The maximum $a$ from each run was recorded and compared to the maximum $a$ of the actual detections 1000 times, as shown in Figure \ref{fig:amax}.
The maximum $a$ of the real high-$q$ non-resonant detections (55.0~AU for the 5:2 and 61.9~AU for the 3:1) are unlikely to be drawn from an underlying uniform distribution.
We find that the maximum $a$ of the near-5:2 population is less than or equal to 55.0 AU (the maximum $a$ of the real OSSOS catalog) only 5.9\% of the time. Moreover, in the bottom panel we see that the detection probability of near-3:1 objects is nearly uniform across the 3:1 resonance. If we account for the fact that the only real near-3:1 detection is on the Sunward side of the resonance, we find a 2.9\% probability that OSSOS would have detected such an asymmetric high-$q$ population if the underlying distribution was uniform within $\pm2$~AU of both resonances.

The OSSOS ensemble of surveys can discriminate between different distributions of near-resonant TNOs, which are known to depend on the specifics of planetary migration, as discussed in Section \ref{sec:sims}.
Using the OSSOS Survey Simulator, {\bf we reject the simple uniform distribution at 2$\sigma$ significance}.
However, the predicted particle distributions resulting from migration are typically more complicated than this simple model.
We test the specific model particle distributions from different planetary migration simulations in the following section.

\subsection{Using the Survey Simulator to Test the Detailed Models} \label{sec:surveysimsetup}

The Survey Simulator software works by randomly drawing an orbit from a file or a parametric distribution, choosing an absolute $H$-magnitude for that drawn object based on an $H$-distribution model, then determining if that drawn object was bright enough and in the right position on the sky to have been detected by the OSSOS ensemble of survey pointings \citep[the Survey Simulator methodology is discussed extensively in previous works, e.g.,][and references therein]{LawlerSurveySimulator}.
In order to make sure that we properly measure the population of high-$q$ resonant dropouts predicted by each model, we compare simulated detections from survey-biased models where there are the same number of simulated detections of resonant objects as real resonant TNO detections in OSSOS.
Because of this scaling and the different detection biases for resonant TNOs, we must carefully preserve the resonant or non-resonant conditions of test particles while we randomize model orbits, another reason this analysis works best with a dynamically classified model.

\begin{figure*}
\begin{center}
\includegraphics[scale=0.44]{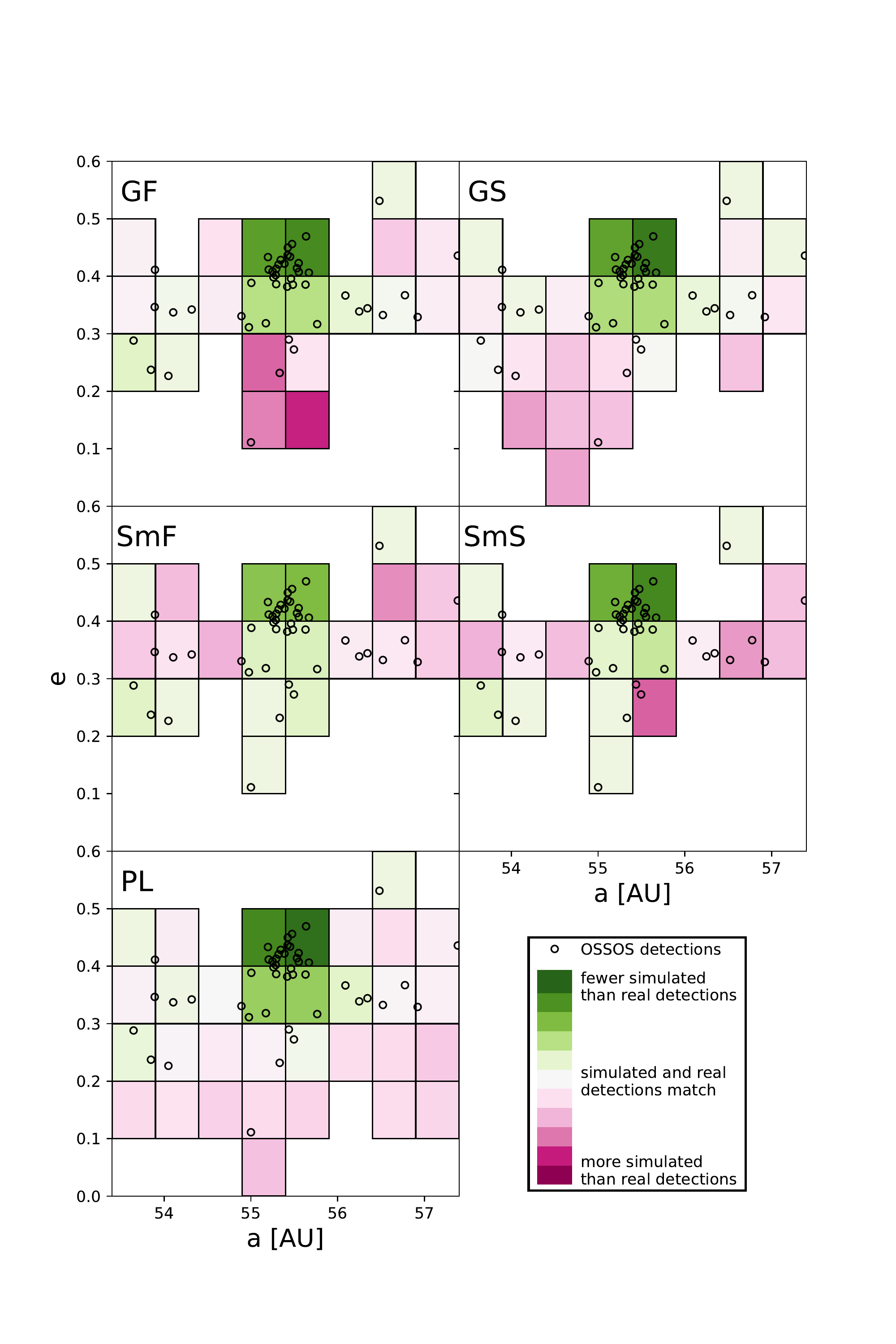} \includegraphics[scale=0.44]{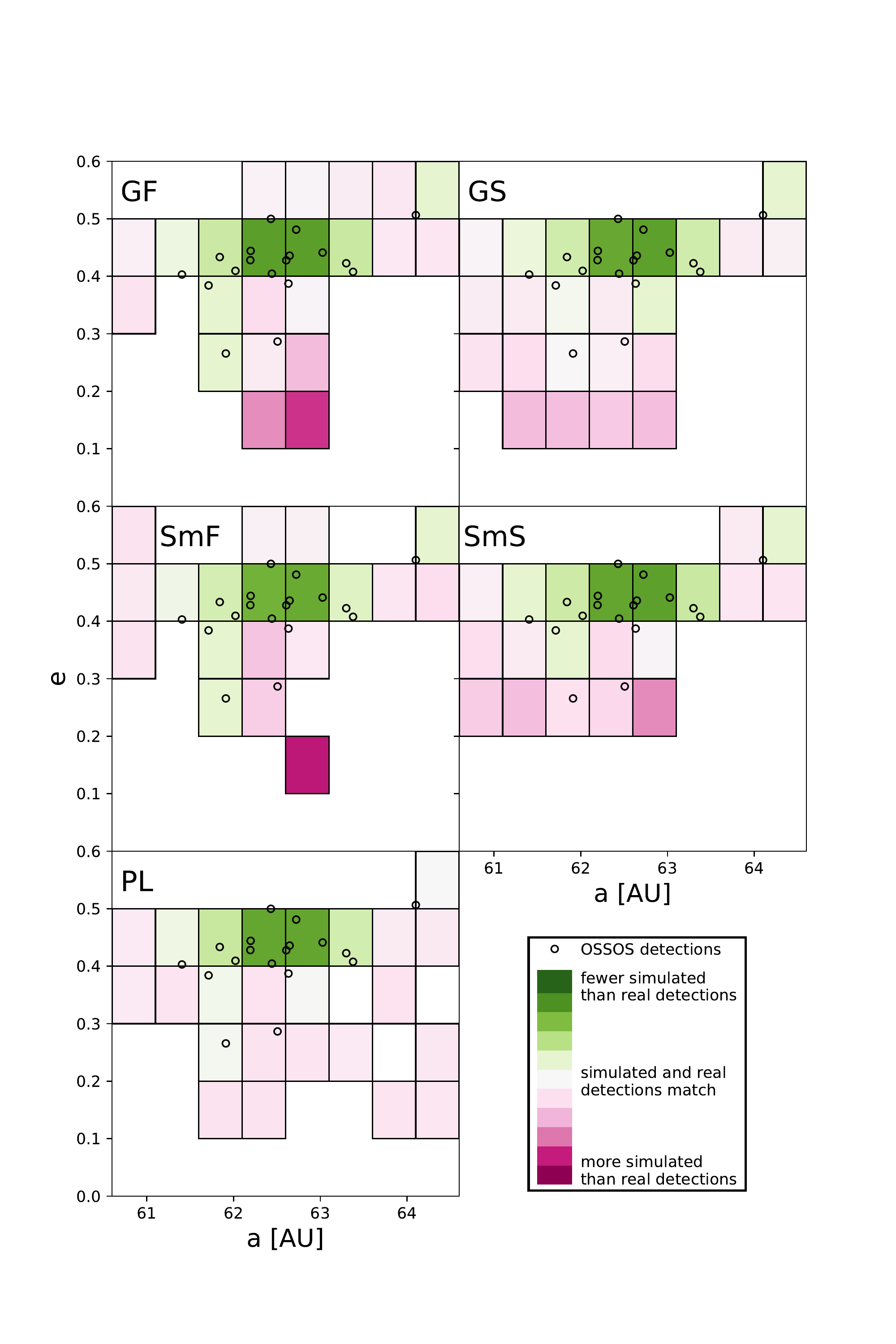}
\caption{
The detectability of objects in each dynamical model compared with the number of real OSSOS detections near the 5:2 (left panels) and 3:1 (right panels) averaged over many Survey Simulator runs. 
See text for the definition of detectability.
Circles show the location of real OSSOS TNOs in $a,e$, and coloured squares show the number of simulated detections from the models compared with real detections in each $a,e$ bin.
Panels are on the same scale for each resonance, with pinker colours showing where the model produces more simulated than real TNO detections, greener colours showing more real TNOs than simulated TNO detections, and grey showing where the match is perfect.
}
\label{fig:gridplot}
\end{center}
\end{figure*}

To produce an orbital distribution from the output of a dynamical simulation as we have done here, orbits are drawn from the model end-state randomly, with each line in the model orbital distribution file being equally likely to be chosen.
If the drawn orbit is classified as non-resonant, the orbits will precess over time, so the angular orbital elements (the argument of pericenter $\omega$, longitude of the ascending node $\Omega$, and mean anomaly $\mathcal{M}$) are all chosen randomly to simulate the distribution over long timescales.
Resonant orbits have a longitude bias: they come to pericenter preferentially at specific  positions on the sky, thus the on-sky pointing positions of the survey are important and can cause non-intuitive detection biases for different resonances \citep[see, e.g., Figure 1 in][]{Gladmanetal2012}.
To preserve this longitude bias, resonant orbits must be randomized in a specific way.  
As part of the classification into resonances \citep[as discussed in detail in][]{Pikeetal2017}, we record the libration amplitude $\Delta\phi_{jk}$ and libration center $\langle\phi_{jk}\rangle$ of each resonant test particle, which limits where the test particle can come to pericenter (and thus be most detectable) relative to Neptune's position.
For the 5:2, all resonant particles have $\langle\phi_{52}\rangle=180^{\circ}$, but the 3:1 has symmetric librators with $\langle\phi_{31}\rangle=180^{\circ}$ and asymmetric librators with $\langle\phi_{31}\rangle\simeq80^{\circ}$ or $280^{\circ}$.
The libration amplitude distribution is different in every simulation, as is the relative population of the different 3:1 resonant islands, but we preserve this distribution inside the Survey Simulator.
For each resonant orbit drawn from the file, following \citet{Gladmanetal2012}, the value of $\phi_{jk}$ is chosen sinusoidally according to the libration amplitude around the resonance center.
The mean anomaly $\mathcal{M}$ and longitude of the ascending node $\Omega$ are then chosen randomly, and the argument of pericenter $\omega$ is chosen to satisfy the resonant condition:
$\phi_{jk}=j\lambda_{\rm TNO}-k\lambda_N-(j-k)(\Omega_{\rm TNO}+\omega_{\rm TNO})$, where $\lambda=\Omega+\omega+\mathcal{M}$ and the subscripts ``N'' and ``TNO'' refer to Neptune and the resonant test particle/TNO, respectively.
In this way, we can randomize orbits and draw many thousands of orbits from a slice of a model that may contain only a few hundred orbits, and still be consistent with the unique resonant dynamics of each model's output.
After the orbits are randomized, an $H$-magnitude is assigned from a literature size distribution \citep{LawlerScattering}, ranging from $4.5<H_r<11$, which covers the range of OSSOS detections in the population we are testing.  The instantaneous distance of the test particle is then used to assign an apparent magnitude, which together with the test particle's instantaneous on-sky position and rate-of-motion is used to determine whether or not this particular particle is detected within the survey characteristics.

We use the Survey Simulator in several different ways to compare the models with the real OSSOS detections near the 5:2 and 3:1 resonances, and to assess how well each model matches reality.

Figure~\ref{fig:gridplot} is an example of the detailed analysis that can be done with dynamically classified models containing many test particles and survey simulator software combined with detections from a well-characterized survey.
We divide the models into $a,e$ bins and then assess the detectability of each bin using the OSSOS Survey Simulator, then compare the survey-biased results with the real detections.
The Survey Simulator is run until 100 objects are detected in each $a,e$ bin (preserving resonance and non-resonance for each drawn model object as described above). 
Detectability is measured by counting how many simulated objects were drawn to reach 100 simulated detections. 
In $a,e$ bins where detectability is low, more simulated objects must be drawn to reach 100 simulated detections than where detectability is high.  
The total detectability is then normalized by requiring the same total number of detections as in OSSOS near the 5:2 and 3:1 resonances. 
The simulated and real detections in each bin can then be directly compared as shown in Figure~\ref{fig:gridplot}, where darker green shows $a,e$ bins where too few simulated detections are produced by the model, darker pink shows bins where too many simulated detections are produced by the model, grey shows good matches between the simulation and reality, and white shows bins with no simulated or real detections.
(The numbers normalized the same for all the panels around the 5:2, and all panels around the 3:1, but the two resonances do not have the same normalization because of the different number of detections.)

Some interesting features are immediately obvious in Figure~\ref{fig:gridplot}.  
Every single model underpredicts the number of TNOs observed at relatively high-$e$ inside the 5:2 and 3:1 resonances (green squares near the center of each panel). 
We speculate that this has something to do with the dynamical excitation of TNOs immediately prior to being captured, and this may be an important feature to attempt to reproduce in future dynamical simulations of Neptune migration.
The numbers of detected TNOs are so small for the high-$q$ (low-$e$) TNOs that it is impossible to evaluate which model here provides the best match in this paramater space.  
We turn to another simulation method for a more direct test.

\begin{deluxetable*}{l|ccc|ccc}
\tabletypesize{\small}
\tablecolumns{9} 
\tablewidth{0pt}
\tablecaption{Resonant and Near-Resonant Detected and Intrinsic Populations \label{tab:summaryTable}}
\tablehead{ & 
\multicolumn{3}{c}{{\bf Normalized to 27 5:2 Resonant TNOs}} & \multicolumn{3}{c}{{\bf Normalized to 11 3:1 Resonant TNOs}} \\
\colhead{Source} & \underline{Intrinsic Populations} & \multicolumn{2}{c}{\underline{Biased by Survey (``Detections'')}} & \underline{Intrinsic Populations} & \multicolumn{2}{c}{\underline{Biased by Survey (``Detections'')}} \\
 & \colhead{high-$q$ near-5:2} & \colhead{high-$q$ near-5:2} & fraction of & \colhead{high-$q$ near-3:1} & \colhead{high-$q$ near-3:1} &  \colhead{fraction of} \\ 
  & \colhead{Sunward/Outward$^a$} & \colhead{Sunward/Outward$^b$} & matching runs$^c$ &  \colhead{Sunward/Outward$^a$} & \colhead{Sunward/Outward$^b$} & \colhead{matching runs$^c$} } 
\startdata
\underline{Models:} \\
KS16 GF & 0.5 / 1.2  & 0.1 / 0.1 & 0.00 & 0.6 / 0.8 & 0.1 / 0.2 & 0.09 \\
KS16 GS & 20.2 / 2.0 & 3.1 / 0.3 & 0.16 & 38.9 / 11.0 &  9.3 / 2.4 & 0.01 \\
KS16 SmF & 0.1 / 0.0 & 0.0 / 0.0 & 0.00 & 1.0 / 1.5 &  0.1 / 0.2 & 0.08 \\
KS16 SmS & 1.1 / 2.3 & 0.1 / 0.4 & 0.00 & 7.7 / 2.2  &  1.0 / 0.3 & 0.24 \\
PL17  & 2.8 / 1.8 & 0.4 / 0.2 & 0.01 & 5.9 / 5.4 &  1.4 / 1.5 & 0.08 \\ \hline
\underline{Real TNOs:}  \\
OSSOS+ & & 3 / 0 & & & 1 / 0   \\ 
\enddata
\tablenotetext{a}{For ease of comparison, the intrinsic population of high-$q$ non-resonant objects in each model has been scaled so that the total number of resonant objects matches the number of OSSOS detections (27 5:2 resonators and 11 3:1 resonators, the near-5:2 populations and near-3:1 populations are scaled independently from each other).}
\tablenotetext{b}{The number of detections is the average number of non-resonant high-$q$ Sunward/Outward detections over 1000 Survey Simulator runs that detected the given number of 5:2 (27) or 3:1 (11) resonators with different random number seeds.  (see Section~\ref{sec:surveysimsetup}).}
\tablenotetext{c}{The fraction of matching runs is the fraction of these 1000 Survey Simulator runs that have \emph{exactly} the same number of Sunward/Outward high-$q$ detections as OSSOS for each resonance.}
\end{deluxetable*}

Table~\ref{tab:summaryTable} shows the result of running each model is run through the Survey Simulator until it produces the same number of resonant detections as discovered by the OSSOS Ensemble.
In order to not be strongly affected by a single random draw, we repeated this 1000 times. 
Table~\ref{tab:summaryTable} gives this average number of high-$q$ near-resonant detections produced by each model when the Survey Simulator produces the same number of resonant detections as in the real OSSOS ensemble.  
The comparison between the fraction of detected Sunward/Outward high-$q$ objects in each simulation is one way to measure the goodness-of-fit of these model to the data.  
We also provide the \emph{intrinsic} populations of high-$q$ non-resonant objects for each model, again scaled so that there are the same number of resonant objects as discovered by OSSOS (obviously the intrinsic population would have to be much larger than the detected population; the scaling is done to provide ease of comparison between models and also to show the relative over-/underpopulation of the Sunward/Outward populations).
As discussed before, because of the complicated survey biases, we do not compare the intrinsic Sunward/Outward fractions directly to the real OSSOS detections, but compare the observationally biased models.
As another way of evaluating the models, we measure the number of Survey Simulator runs where the exact OSSOS Sunward/Outward number of detections are matched exactly for each resonance.

Looking at the 5:2 portion of Table~\ref{tab:summaryTable}, it is clear that the KS16 Grainy Slow simulation provides the best match to the OSSOS data.  
The average number of Sunward and Outward high-$q$ objects detected is the closest to the real OSSOS detections, and the fraction of runs where exactly 3 Sunward and 0 Outward high-$q$ objects are detected is the highest.
The PL17 model is the second best match for the 5:2. 
Coincidentally, the PL17 intrinsic Sunward/Outward population ratios are actually closest to the OSSOS detections, but when biased by the Survey Simulator, drop to much lower values than the real OSSOS detections, with only 1\% of the Survey Simulator runs producing exactly 3 high-$q$ Sunward and 0 Outward detections along with the 27 5:2 resonant detections.  
Interestingly, the KS16 Grainy Fast, Smooth Fast, and Smooth Slow produce zero or near-zero runs where there is an exact match to the OSSOS high-$q$ detections.  
This is because the resonant population is much higher relative to the non-resonant in these simulations than in the Grainy Slow and PL17 models, so there are simply not enough non-resonant detections by the time the Survey Simulator has produced 27 5:2 resonant detections in each of these three models.
For the near-3:1 objects, the KS16 Grainy Slow simulation provides too many high-$q$ detections both Sunward and Outward, while the KS16 Smooth Slow migration produces the best fit, though we note that with only one high-$q$ OSSOS detection to compare here, this analysis holds little statistical weight and is mainly qualitative.

\subsection{Grainy Slow Migration Likely Provides the Best Match} \label{sec:gsbest}

There are many possible ways to test dynamical models against the OSSOS data; we remind the reader that here we are focusing only on the high-$q$ resonant dropouts because their ``fossilized'' dynamically detached state means that particles that manage to get into this parameter space will stay on a very similar orbit on Gyr timescales.  
Comparison of orbital element, libration amplitude distributions, and Kozai fractions within the resonances has already been done for the PL17 simulation in that work.
Comparison between the models and OSSOS resonant data will be in upcoming OSSOS team papers specifically focusing on the discoveries and orbital structure of the 5:2 and 3:1 resonances.

The number of high-$q$ non-resonant detections produced by the KS16 Grainy Slow model when matching the OSSOS resonant detections agrees well for the 5:2 (Table~\ref{tab:summaryTable}).  
However, when comparing the near-3:1 simulated detections, the number predicted by the KS16 GS model is far too large compared with OSSOS, and the PL17 and KS16~SmS models provide a better match.  
But with only one high-$q$ detection near the 3:1, it is extremely difficult to say anything significant about these comparisons.

\begin{figure*}
\begin{center}
\includegraphics[scale=0.7]{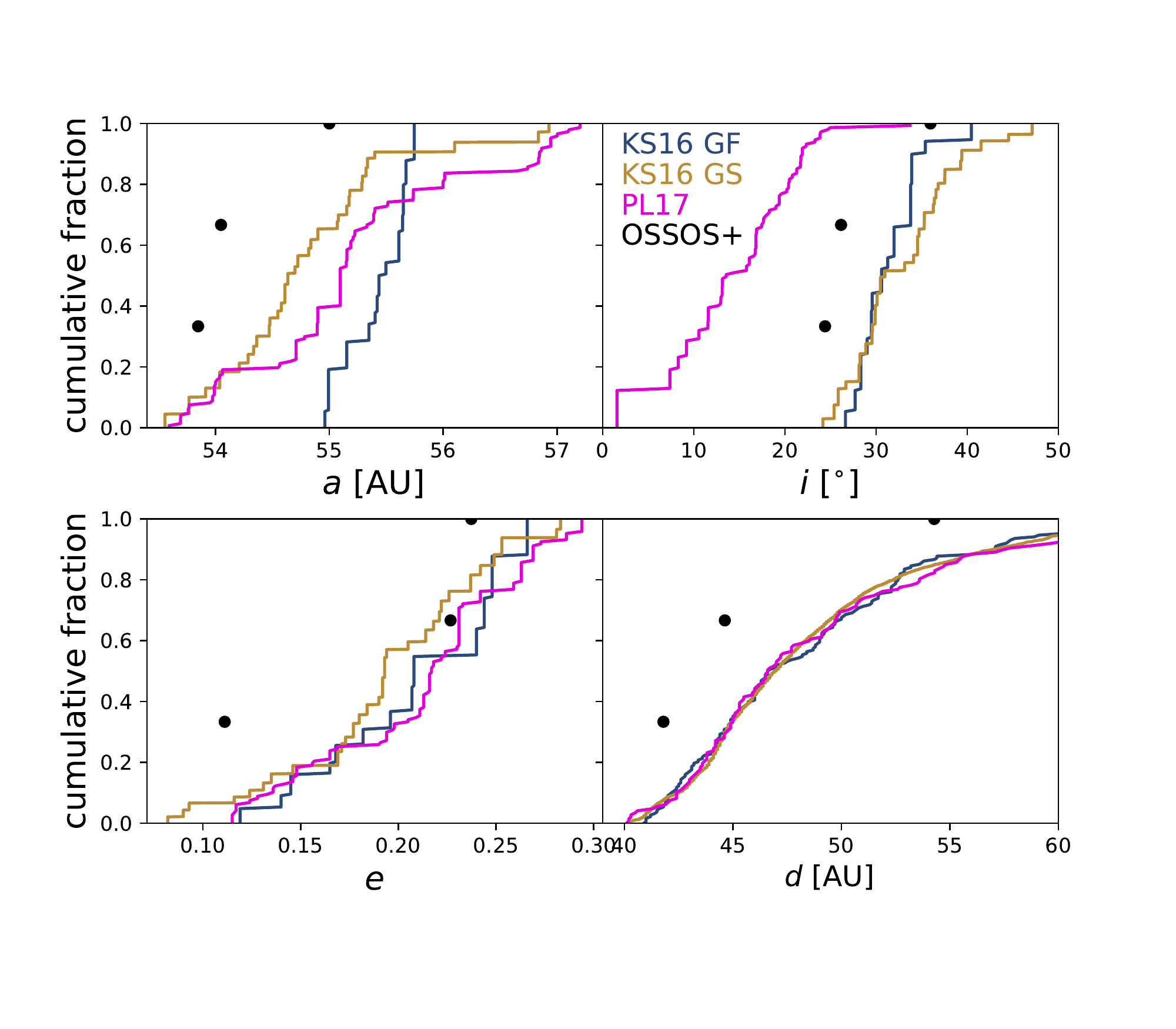}
\caption{
Comparison between the three high-$q$ OSSOS detections (black points) and properties of simulated non-resonant high-$q$ detections from three of the models that have a significant number of high-$q$ non-resonant objects near the 5:2.  
Because of the very small numbers of real OSSOS detections in this parameter space we present this mainly as a guide for future comparisons; see discussion in the Appendix.
Only KS16 Grainy Fast is statistically ruled out by this analysis; it does not have a wide enough range of $a$ or $i$ to encompass the real OSSOS detections.
}
\label{fig:cumuplot}
\end{center}
\end{figure*}

Another way to compare the simulated detections is provided in Figure~\ref{fig:cumuplot}, where the three high-$q$ near-5:2 OSSOS detections are compared with cumulative distributions of the simulated model detections of high-$q$ objects for three of the models that have the most high-$q$ objects near the 5:2.
The KS16~GS and PL17 models both easily provide more Sunward than Outward of the 5:2 resonance, while the KS16~GF provides detections only in a very narrow range of $a$.  
The PL17 model provides overall lower inclinations in the high-$q$ population than either of the KS16~GS or GF models, but it is unclear from the small number of detections which provides a better match to the OSSOS data.
Only the KS16~GF model can be ruled out: Due to the fast migration timescale, resonant dropouts appear to only happen very close to the end position of the resonance, keeping a very narrow $a$ distribution that does not encompass the range of OSSOS detections.
Using the Anderson-Darling test \citep{AndersonDarling}, neither the PL17 or KS16~GS models are formally rejectable by the data presented here.
The eccentricity distribution between the three models is remarkably similar, as is the distribution of distances where objects are detected.
We reiterate that with so few detections, Figure~\ref{fig:cumuplot} does little more than provide guidelines for future work which will hopefully provide far more high-$q$ TNO detections (see Appendix).

\section{Discussion} 

\subsection{Comparison with the MPC Database} \label{sec:mpc}

\begin{figure*}
\begin{center}
\includegraphics[scale=0.7]{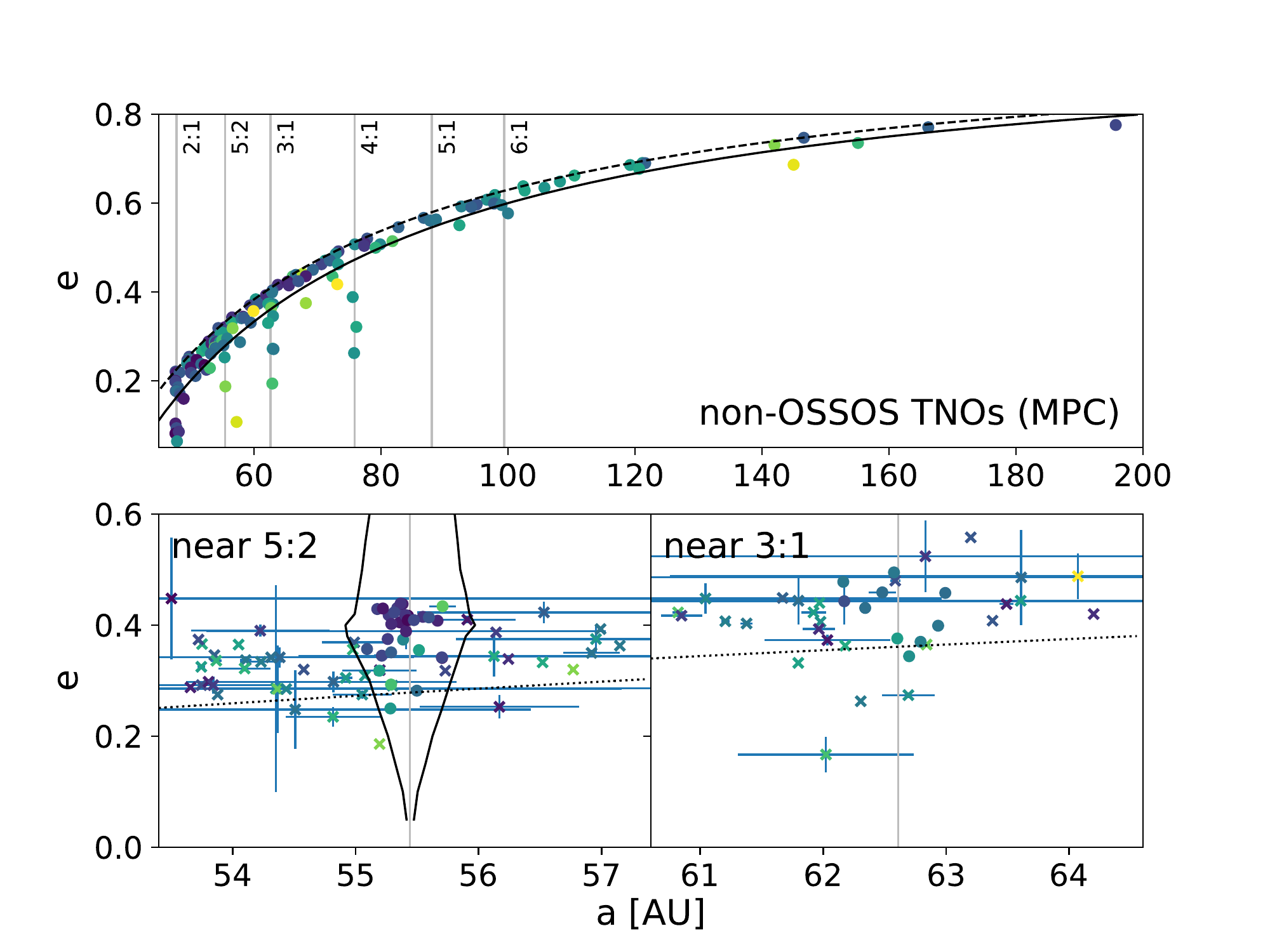}
\caption{
Orbital elements of TNOs from the Minor Planet Center Database as of 2018 December with $q>37$~AU, $a>47.6$~AU (Outward from the location of the 2:1 resonance), and an MPC uncertainty parameter of $<6$, excluding those detected by the OSSOS ensemble.
Color scale for inclination is as in Figure~\ref{fig:detached}.
In top panel, the solid and dashed black lines show $q=40$ and 37~AU, respectively.
In bottom panels, dotted line shows $q=40$~AU.
The top panel can be compared to the real OSSOS detections in Figure~\ref{fig:detached}, and the bottom two panels can be compared with the real OSSOS detections in Figure~\ref{fig:ossos}.
In the bottom panels, orbital classification has been performed in the same manner as for OSSOS, circles show TNOs that librate in the 5:2 and 3:1 resonances.
Most of these TNOs have orbital uncertainties that are so large that they are not useful for the sort of analysis performed here; additionally, observation biases for these TNOs are unknown.
}
\label{fig:mpc}
\end{center}
\end{figure*}

Our analysis has focused only on TNOs detected in well-characterized surveys, here we also compare with all the detected TNOs currently listed in the Minor Planet Center (MPC) Database (as of 2018 December).
There are more TNOs listed in the MPC Database than the OSSOS ensemble, but they come from a myriad of surveys with unreported observation biases, so direct comparison between models and detections is impossible without large uncertainty.
Figure~\ref{fig:mpc} shows all TNOs in the MPC Database with $a$ larger than the 2:1 resonance center and $q>37$~AU.
Due to the very large uncertainties on most of the data, we did not perform dynamical classification to determine resonance occupation or scattering behavior.  

The lower two panels of Figure~\ref{fig:mpc} show the near-5:2 and near-3:1 TNOs, which can be compared to the OSSOS detections plotted in Figure~\ref{fig:ossos}.  
The pointings of the OSSOS survey were specifically chosen to optimize detection of $n$:2 resonators \citep{Bannisteretal2016}, so there are many more 5:2 resonant detections in OSSOS than the number of MPC TNO detections inside the \citet{Malhotraetal2018} 5:2 stability limits.
Other than a lack of dynamically excited 5:2 resonators (almost certainly an observation bias effect), the MPC database detections are completely consistent with the OSSOS detections.
We do not include these extra TNOs in our analysis because the observation biases are completely unknown, thus the detailed observational bias analysis performed above is not possible for this sample.

\subsection{The Dynamics of Resonant Dropouts} \label{sec:dropouts}

The orbital distribution of these fossilized resonant dropouts is a window into the construction of our Solar System's architecture.
The interplay of Kozai oscillations within resonances and Neptune's migration appears to explain much of the resonant and non-resonant Kuiper belt dynamical structure that we observe.
Grainy migration gives us information not only on Neptune's migration, but on the number of large ($\sim1000$~km) planetesimals that formed before Neptune's migration occurred. 
The analysis performed here provides additional support for the idea that there were hundreds of such large planetesimals present by the time of Neptune's migration, as per \citet{ShannonDawson2018}, and supports the theory of \citet{NesvornyVokrouhlicky2016}.

While grainy slow migration appears to provide a satisfactory explanation for the high-$q$ TNOs at $a<100$~AU, it is not clear if this could also be applied to TNOs at much larger distances.  
Recent simulations by \citet{Yuetal2018} show that at any given moment, for TNOs with $30<a<100$~AU, there are of order the same number of scattering TNOs as transiently resonant TNOs.
Transient sticks to mean-motion resonances by the unstable scattering population provides a nice explanation for the very large populations present in very distant mean-motion resonances \citep{Gladmanetal2012,Pikeetal2015,Bannisteretal2017,Volketal2018,Holmanetal2018}.
\citet{Pikeetal2015} performs a detailed analysis of TNOs in and around the 5:1 resonance, and show there is a non-resonant near-5:1 TNO in a high-$q$, long-term stable orbit Sunward of the resonance. 
At $a=88$~AU, the 5:1 is not expected to have any primordial TNO population, and must be entirely populated by resonant sticking from the scattering population in any of the Neptune migration models we discuss in this work.
Just after the Nice-model planet scattering event, the scattering population would have been much larger; the scattering population today is likely $<1\%$ of its original population \citep{DuncanLevison1997}.
Resonant sticking by this much larger scattering population at early times, combined with grainy Neptune migration, could explain high-$q$ TNOs at much larger semimajor axes.  

The two highest-$q$ TNOs discovered to date have pericenter distance more than 25~AU larger than the next highest-$q$ TNOs, possibly implying a different formation mechanism than other high-$q$ TNOs: Sedna has $a=507$~AU and $q=76$~AU, and 2012~VP$_{113}$ has $a=266$~AU and $q=80$~AU.
For the same resonant sticking, Kozai $q$-lifting, and resonant dropout narrative to work for these two extreme TNOs, they must have been caught into very large-$a$ resonances.  
Sedna is near the semimajor axis that would correspond to its 69:1 resonance, and 2012~VP$_{113}$ is close to the 105:4 resonance.  
Initially it seems a bit ridiculous to think that these extremely high-order, distant resonances could possibly affect TNOs, but resonance occupation by real TNOs has been shown in such high-order resonances as, e.g., the 27:4 and the 35:8 at $a=108$~AU and 81~AU, respectively \citep{Bannisteretal2018}.
Additionally, simulations by \citet{Gallardoetal2012} found resonance sticking with pericenter lifting out to the outermost semimajor axes they tested at 500~AU, and test particles within these simulations had their pericenters raised up to $q\simeq70$~AU.
However, they show that for $q$ to change dramatically, inclinations generally must be high ($i\gtrsim40^{\circ}$) when particles are first captured into resonance, and these TNOs' inclinations do not appear to be high enough for this explanation to be compelling: Sedna has $i=11^{\circ}$ and 2012~VP$_{113}$ has $i=24^{\circ}$.
There are many suggested explanations for distant, high-$q$ TNOs in the literature, but using Neptune's migration in combination with resonance sticking is an intriguing and much simpler possible explanation for all but the highest-$q$ TNOs and thus merits further exploration in detailed simulations.
Repeating Neptune migration simulations with many more scattering test particles at large-$a$ and carefully classifying and analyzing test particle behavior near the distant resonances is a highly fruitful route to measuring measure the effectiveness of this population mechanism and may provide an answer to source of many of the fossilized high-$q$ objects.

\section{Summary and Conclusions}

The current observations of the outer Solar System show that the distribution of high-$q$ particles near the 5:2 and 3:1 resonances are not uniform.
The analysis here has shown that more high-$q$ TNOs on the Sunward side of these two resonances is not an observational bias effect, and the structure is likely due to resonant dropouts during Neptune migration.
The distribution of these near-resonant particles provides a powerful discriminating tool for assessing how well different models of planetary migration reproduce the outer Solar System.
The current four characterized near-resonant TNO detections, all on the Sunward side of nearby resonances, provide a statistically significant rejection of a uniform distribution, and indicate that Grainy Slow Neptune migration \citep[as in][]{KaibSheppard2016} provides the closest match to the observed near-resonant TNO distribution.

Out of the five Neptune migration simulations analyzed, the simulation which used {\bf grainy Neptune migration with slower timescales} and a scattering jump that keeps Neptune at $e<0.1$ best reproduces the relative number and distribution of near-resonant high-$q$ TNOs observed by OSSOS, but there simply aren't yet enough high-$q$ TNOs to provide statistically robust support for any of the models we tested above the others.
We hope this paper has provided a framework to rigorously test future observational datasets against upcoming dynamical models.

The Large Synoptic Survey Telescope is expected to detect hundreds of new TNOs because of its incredible sky coverage and cadence.
The LSST main survey of 18,000 square degrees has a single-image limiting magnitude of $m_r=24.5$ \citep{Ivezicetal2008,Jonesetal2016}, shallower than much of OSSOS. 
It will thus provide comparatively few high-q TNOs, due to the steep TNO size distribution. 
However, a Deep Drilling Survey of only a few suitably sited pointings, with a comparatively sparse cadence, may also be able to differentiate between these models.  
The most powerful way to test migration models will be through a deeper targeted TNO survey, such as the proposed survey we modeled for Subaru HSC (see Appendix), which would detect enough high-$q$ TNOs to distinguish between these five migration models at high statistical significance.

Additionally, larger-scale Neptune migration simulations with more scattering particles can test the effectiveness of resonant sticking by scattering TNOs during Neptune's migration phase for producing very high-$q$ TNOs at large semimajor axes, such as Sedna.  
Due to the extreme biases against observing these extremely high-$q$ TNOs and the low number of detections, the best way to understand their population is through dynamical modelling in combination with a Survey Simulator to fully account for these severe observation biases.

By reconstructing the details of Neptune's migration through detailed study of the Kuiper Belt's orbital structure, we learn about possible additional, now-ejected giant planets \citep[as suggested by the Jumping Jupiter model;][]{Nesvorny2015b}, and the size and number of planetesimals that had formed at this distance, since grainy migration requires a large number of relatively large ($\sim$1000~km) planetesimals \citep{NesvornyVokrouhlicky2016}.
The large OSSOS dataset and Survey Simulator, which are both publicly available, provide the most powerful way to test these and future dynamical evolution models of the Solar System. 
Using a Survey Simulator is the best way to take into account the complicated observational biases in the outer Solar System, particularly in the high-$q$ and resonant TNO populations, and we hope that this will become the standard for testing the validity of dynamical models.

\acknowledgments

The authors acknowledge the sacred nature of Maunakea and appreciate the opportunity to observe from the mountain. CFHT is operated by the National Research Council (NRC) of Canada, the Institute National des Sciences de l'Universe of the Centre National de la Recherche Scientifique (CNRS) of France, and the University of Hawaii, with OSSOS receiving additional access due to contributions from the Institute of Astronomy and Astrophysics, Academia Sinica, Taiwan. Data are produced and hosted at the Canadian Astronomy Data Centre; processing and analysis were performed using computing and storage capacity provided by the Canadian Advanced Network For Astronomy Research (CANFAR).
This research has made use of data and/or services provided by the International Astronomical Union's Minor Planet Center.

The authors thank R.\ Brasser for providing the output of his simulation for this analysis, C.\ Shankman for legacy code and helpful conversations, H.\ Ngo for useful conversations, and an anonymous referee for suggestions which improved the focus of the paper.
S.M.L.\ gratefully acknowledges support from the NRC-Canada Plaskett Fellowship. 
M.T.B. appreciates support from UK STFC grant ST/P0003094/1.
K.V. acknowledges support from NASA grants NNX15AH59G and NNX14AG93G and NSF grant AST-1824869.

\facilities{CFHT, CANFAR} 
\software{matplotlib \citep{Hunter2007}, 
scipy \citep{Jonesetal2001},
Mercury \citep{Chambers2001}, 
SWIFT \citep{LevisonDuncan1994},
OSSOS Survey Simulator \citep{LawlerSurveySimulator,Petitetal2018}}

\appendix
\section{Future Prospects for TNO Discovery Surveys} \label{sec:futuresurvey}

Determining the significance of the detections Sunward and Outward of the resonances requires that the detections are from a characterized survey with known selection effects.
The OSSOS ensemble of surveys provides a powerful tool for testing different models of the outer Solar System.
These surveys show that there is a statistically significant population found Sunward of the resonances.
However, this survey has few detections of these near-resonant objects, which results in a large uncertainty in the statistics and the population size.
We determine here the effectiveness of a deeper survey in comparing the Sunward and Outward populations in these models.
We use as an example a survey strategy performed by Hyper Suprime-Cam (HSC) on Subaru Telescope with the same layout and field coverage as the OSSOS survey.

The Hyper Suprime-Cam Subaru Strategic Program (HSC-SSP) included a TNO search, but the survey cadence was not well suited to TNO identification \citep{chen2018}.
However, the detections from this survey confirm the powerful nature of HSC in a TNO search survey.
The HSC-SSP utilized 180 second exposures in $g$, $r$, $i$, $z$, and $Y$ bands.
In $r$, the optimal band for TNO searches, the limiting magnitude of the survey was $m_r=25.0-25.5$ \citep{chen2018}.
For a TNO search, exposures up to 300 seconds can be used to maximize depth while keeping trailing loss minimal (for sub-arcsecond seeing).
With an increased exposure time of 300 seconds, we expect HSC would deliver a survey limiting magnitude of $m_r=25.5-26.0$.
Because of the large area, the HSC-SSP was also conducted in a range of sky conditions, including poorer seeing and transparency than the OSSOS discovery survey.
A targeted TNO search would require better conditions, and we are confident that those survey blocks would consistently deliver detections to a limiting magnitude $m_r=26.0$ or fainter.
Subaru is the only 8m-class telescope that has a wide-field camera: HSC has a 90 arcminute diameter field-of-view. 
HSC would also provide a faint limiting magnitude in average observing conditions; OSSOS was able to attain $m_r\simeq25.2$ in the best seeing conditions with the 3.6m aperture CFHT. 
HSC can provide the optimal system for pushing the frontiers of TNO science.

\begin{figure}
\begin{center}
\includegraphics[scale=0.6]{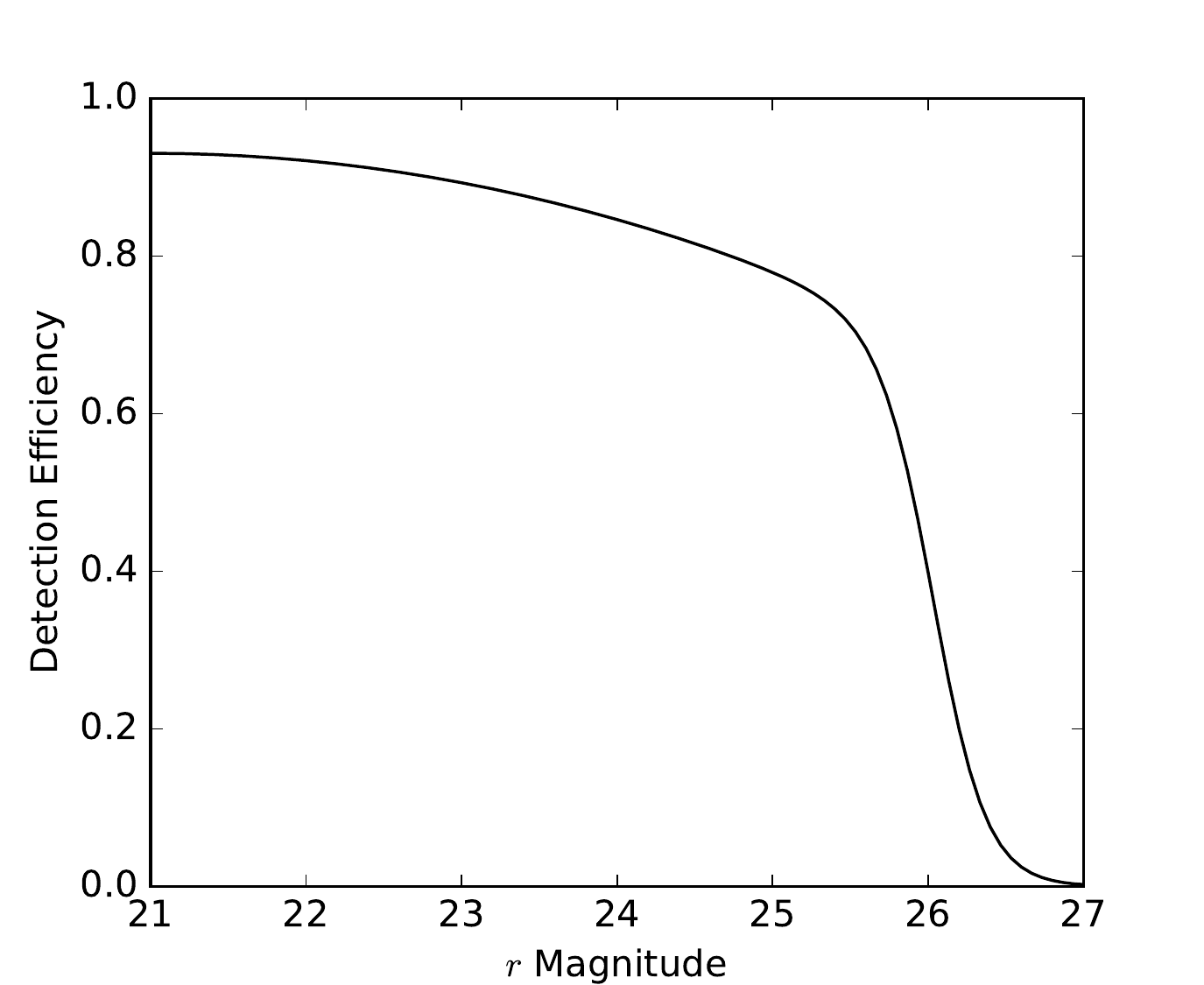}
\caption{
The assumed detection efficiency of the simulated HSC survey blocks.
The parameterization is described in \citet{Bannisteretal2016,Bannisteretal2018}, and is modeled after the deepest OSSOS block sensitivity (D-block, $m_r\simeq25.2$), but with a fainter limiting magnitude.  The survey detection limit is $m_r=26$, where the detection efficiency drops to $\sim$40\%. 
}
\label{eff}
\end{center}
\end{figure}

We use the known characteristics of HSC to define a simple characterization of a discovery survey on Subaru.
To simplify the comparison between with current surveys, we use the same field pointings and areas as the OSSOS surveys \citep[described by][]{Bannisteretal2016,Bannisteretal2018}.
We expect a similar detection efficiency to OSSOS because of the similar chip gaps and expected moving object detection pipeline method.
Here this is parameterized using the same characteristics as the OSSOS D-block efficiency, but with a fainter dropoff magnitude ($m_r=26.05$), see Figure \ref{eff}.
We use a magnitude limit of $m_r=26.0$ based on 300 second exposures on HSC as discussed above, and characterize the detection efficiency drop based on the OSSOS efficiency curves \citep{Bannisteretal2016}, as this is closely related to the survey cadence and search method.
If the discovery survey is conducted in the same cadence as the OSSOS survey, the same moving object detection pipeline \citep[described in][]{Petitetal2004} will provide the best TNO identification tool.
This characterization provides a tool for us to test the utility of a deeper HSC survey.

We use the KS16 and PL17 TNO distribution models as sample populations for this theoretical survey.
To determine the absolute number of detections, a careful scaling of the model is required to ensure that it can be compared to the actual Solar System.
For each model, we scaled the number of resonant objects inside the 5:2 and 3:1 and the nearby objects so that each resonant population matches the population estimate found by the Canada-France Ecliptic Plane Survey \citep[CFEPS;][]{Gladmanetal2012}\footnote{We note that \citet{Volketal2016} provides a more recent estimate of the 5:2 population using CFEPS and the first two blocks of OSSOS, but due to the way that the population was modelled, the estimate given in that work is a lower limit. A more precise population estimate for the 5:2 based on the entire OSSOS survey is forthcoming.}.
The 5:2 resonators are reported to have 12,000$^{+15,000}_{-8,000}$ members with $H_r<8.46$ \citep[here $g$ magnitudes have been transposed to $r$ using $g-r=0.7$, shown to be consistent with the colors of dynamically excited TNOs;][]{Shankmanetal2016}.
The 3:1 resonators have 4,000$^{+9,000}_{-3,000}$ members in the same size range, according to CFEPS.
At pericenter, the faintest detectable objects will have $H_r$=11.6 and 11.1 for the 5:2 and 3:1 resonances, respectively.
However, the objects of interest are always $q>40$, so we only include objects above that detectability limit, $H_r<$10.
Assuming the model population $H$-magnitude slope of $\alpha=0.9$ extends to $H_r=10$, we expect 68,000 5:2 objects and 23,000 3:1 resonators with $H_r<10$.
We duplicated the non-resonant model objects with $q>40$ and within $\pm2$~AU of the resonance, retaining the $a$, $e$, and $i$ and randomizing the angular orbital elements in order to produce this large model.
For example, \citet{Pikeetal2017} identified 337 5:2 resonators and 77 3:1 resonators in the \citet{BrasserMorbidelli2013} simulation, so the number of particles near these resonances are scaled by factors of 200 and 300, respectively, to a total of 67,400 and 23,100 particles near the resonances.
(See Table \ref{tab:deepOSSOS} for the size of each input near-resonant model.)

We then used the OSSOS Survey Simulator to `observe' these scaled models with our simulated HSC survey, and determined the $a$-distribution of detected $q>40$~AU objects near the 5:2 and 3:1 resonances.
We repeated this process 100 times for each model and resonance to obtain a large sample of high-$q$ $a$-distributions detected by the surveys.
A large forward-biased comparison model was created for each population by combining the simulated detections from all 100 sets.
Each of the survey simulator runs (100$\times$5 models$\times$2 resonances) was compared to the 5 large comparison models for that resonance using the Anderson-Darling statistical test \citep[AD;][]{AndersonDarling}.
The significance of the AD statistic was determined using a bootstrapping method.
Then the percentage of the 100 models which are \emph{not} rejected at 3$\sigma$ significance by the AD statistic was determined.
Table \ref{tab:deepOSSOS} summarizes these results: a measurement of whether a deeper survey would be able to discriminate between the different proposed intrinsic models.

\begin{deluxetable}{ll|cc|ccccc}
\tabletypesize{\small}
\tablecolumns{9} 
\tablewidth{0pt}
\tablecaption{Expected Rejectability Results from a Survey to $m_r\le26$ \label{tab:deepOSSOS}}
\tablehead{ & \colhead{Model} & \colhead{\# High-$q$ Sim.\ Detections} & \colhead{Number of Particles} & \multicolumn{5}{c}{$a$ Distribution: \% Not Rejectable (3$\sigma$)} \\
 & & \colhead{Mean$\pm$Standard Deviation} & \colhead{$H_r<10$} & \colhead{KS16 GF} & \colhead{KS16 GS} & \colhead{KS16 SmF} & \colhead{KS16 SmS} & \colhead{PL17} } 
\startdata
 & KS16 GF & 36 $\pm$ 6 & 53,622 & 100\% & 26\% & 73\% & 16\% & 3\% \\
 & KS16 GS & 60 $\pm$ 8 & 205,700 & 41\% & 100\% & 64\% & 50\% &2\%  \\
{\bf 5:2} & KS16 SmF & 30 $\pm$ 6 & 36,540 & 39\% & 35\% & 94\% & 33\% & 8\% \\
 & KS16 SmS & 26 $\pm$ 5 & 72,216 & 42\% & 91\% & 63\% & 100\% & 32\% \\
 & PL17 & 15 $\pm$ 4 & 67,400 & 98\% & 100\% & 100\% & 99\% & 98\% \\ \hline
  & KS16 GF & 18 $\pm$ 5 & 48,545 & 99\% &45\% & 45\% & 63\%  & 2\% \\
 & KS16 GS & 32 $\pm$ 6 & 203,435 & 18\% & 100\% & 18\% & 15\% & 1\% \\
{\bf 3:1} & KS16 SmF & 17 $\pm$ 4 & 45,990 & 84\% & 17\% & 100\% & 83\% & 2\% \\
 & KS16 SmS & 16 $\pm$ 4 & 64,400 & 49\% & 9\% & 31\% & 100\% & 1\% \\
 & PL17 & 3 $\pm$ 2 & 23,100 & 87\% & 87\% & 87\% & 86\% & 87\% \\ 
\enddata
\tablecomments{The acceptability of the models based on the simulated detections.  The $a$-distributions are compared using the AD statistical test.  The 100 sets of simulated detections (left column) are compared to the large combined comparison models (top line).  The percentage of sets of simulated detections for which the AD statistic finds that they are consistent with being drawn from the reported model is reported at 3$\sigma$ significance.  As expected, the simulated detections are consistent with being drawn from their own model.  However, the majority of models are statistically inconsistent with each other, and could be differentiated by this survey.
}
\end{deluxetable}

The comparison between the different Survey Simulator runs on the models and the larger biased model of each simulation type shows that this deeper survey would successfully provide stronger conclusions about the underlying TNO distribution.
There are several main conclusions we draw from the AD statistic results in Table \ref{tab:deepOSSOS}.
Importantly, the individual survey simulator runs are consistent with their biased input model with the statistical significance expected, $\sim$99\% of the models are not rejectable at $3\sigma$ confidence.
When we compare the individual survey simulator runs with the biased larger models from other simulations, it is clear that while some models can't be cleanly differentiated, some are clearly distinct.
If the PL17 model is representative of the intrinsic distribution, we will be able to rule out all KS model varieties.
If the real underlying distribution resembles any of the KS16 models, it appears that we will have less statistical significance in rejecting the PL17 model.
However, the significantly fewer detections this model produces is both the reason that these models are not rejected and another criteria we can use to assess rejectability.  
As a result, a future deeper survey should conclusively determine whether Neptune's migration resembles the KS16 or PL17 models.

Comparing the different KS16 models shows that we may be able to differentiate between some of the specific parameters of Neptune migration.
We find that, as expected, models which share components of the same Neptune migration are more similar.
For example, the Smooth Fast (SmF) for both the 5:2 and 3:1 near-resonant particles is similar to the Grainy Fast (GF) and Smooth Slow (SmS) simulations, but may be distinguishable from the Grainy Slow (GS) distribution.
Looking at the combination of the 5:2 and 3:1 near-resonant detections will be key to understanding the underlying high-$q$ object distribution.

Here we have assumed a single slope size distribution.
Assuming that the high-$q$ TNOs have the same size distribution as other dynamically excited populations, we note that this survey would detect some high-$q$ objects beyond the size distribution transition discovered in several dynamically excited populations \citep{Fraseretal2014,Alexandersenetal2016,LawlerScattering}.
However, the majority of large-$H$ objects in our model are not detected because of their reduced brightness, so this effect should not have a significant impact on the results computed here.
As a worst-case scenario, we calculate the effect of a size distribution at $H_r=8.5$ from $\alpha=0.9$ to $\alpha=0.5$ on the KS16 SmF model, which had the most large-$H$ detections.
We find that 60\% of the detections would be retained for this model, resulting in 10 near-resonant detections.
When combined with the $\sim$16 detections of near-5:2 objects, these detections would still provide a powerful tool for measuring the underlying near-resonant TNO distribution.

A TNO discovery survey by HSC would be able to differentiate between the \citet{KaibSheppard2016} and \citet{PikeLawler2017} models at $2-3\sigma$ significance.
Some of the simulations, particularly the KS16 Smooth migrations, also produce smaller numbers of near-resonant populations.
In addition to future TNO searches, it will be critical to increase dynamical model resolution near the resonances by increasing the number if test particles used.
A combination of additional simulations and deeper TNO surveys is required to understand the distribution and history of high-$q$ TNOs.

\bibliographystyle{yahapj}

 \newcommand{\noop}[1]{}

\end{document}